\documentclass[11pt]{article}

\PassOptionsToPackage{table}{xcolor}
\usepackage[final]{acl}

\usepackage{times}
\usepackage{latexsym}
\usepackage[T1]{fontenc}
\usepackage[utf8]{inputenc}
\usepackage{microtype}
\usepackage{graphicx}

\usepackage{amsmath}
\usepackage{booktabs}
\usepackage{pifont}
\usepackage[most]{tcolorbox}
\tcbuselibrary{breakable,skins}
\usepackage{fvextra}
\usepackage{multirow}
\usepackage{listings}
\usepackage{inconsolata}
\usepackage{subcaption}
\usepackage{framed}
\usepackage{threeparttable}
\usepackage{placeins}

\let\oldding\ding
\renewcommand{\ding}[1]{%
  \ifnum#1=51
    \textcolor{green!60!black}{\oldding{#1}}%
  \else\ifnum#1=55
    \textcolor{red!70!black}{\oldding{#1}}%
  \else
    \oldding{#1}%
  \fi\fi
}

\newtcolorbox{finding}{
    colback=gray!20,
    colframe=white,
    boxrule=0pt,
    arc=3pt,
    left=4pt,
    right=4pt,
    top=3pt,
    bottom=3pt,
    boxsep=0pt,
    shadow={0pt}{-0.5pt}{0pt}{black!30}
}

\definecolor{grammarbg}{RGB}{245,245,245}
\newtcolorbox{prompt}{
  enhanced,
  breakable,
  ignore nobreak=false,
  colback=grammarbg,
  colframe=black,
  boxrule=1.2pt,
  arc=8pt,
  outer arc=8pt,
  left=5pt, right=5pt, top=5pt, bottom=5pt,
  before skip=4pt,
  after skip=4pt,
  boxsep=0pt
}

\DefineVerbatimEnvironment{PromptText}{Verbatim}{
  breaklines=true,
  breakanywhere=false,
  breaksymbolleft={},
  breaksymbolright={},
  breakindent=0pt,
  fontsize=\fontsize{10.5}{12.6}\selectfont,
  formatcom=\normalfont
}

\DeclareRobustCommand{\benchname}{{\upshape\scshape CodeSpecBench}}

\definecolor{lbcolor}{rgb}{0.95,0.95,0.95}
\definecolor{dkgreen}{rgb}{0,0.6,0}
\definecolor{mauve}{rgb}{0.58,0,0.82}
\lstnewenvironment{Python}{
  \lstset{frame=tb,
  language=Python,
  backgroundcolor=\color{white},
  aboveskip=1mm,
  numbersep=2pt,
  xleftmargin=10pt, %
  belowskip=3mm,
  showstringspaces=false,
  columns=flexible,
  basicstyle={\scriptsize\fontfamily{zi4}\selectfont\color{black}},
  numbers=left,
  numberstyle=\tiny\color{mauve},
  keywordstyle=\color{blue},
  commentstyle=\color{mauve},
  stringstyle=\color{dkgreen},
  breaklines=true,
  breakatwhitespace=true,
  tabsize=3,
  upquote=true}
}{}

\definecolor{deepred}{rgb}{0.6,0,0}
\definecolor{deepblue}{rgb}{0,0,0.5}

\title{\benchname: Benchmarking LLMs for Executable \\ Behavioral Specification Generation}

\author{%
\textbf{Zaoyu Chen}$^{1}$\thanks{Equal Contribution.},
\textbf{Jianbo Dai}$^{2*}$,
\textbf{Boyu Zhu}$^{3}$,
\textbf{Jingdong Wang}$^{4}$,
\textbf{Huiming Wang}$^{5}$,  \\
\textbf{Xin Xu}$^{6}$,
\textbf{Haoyang Yuan}$^{7}$,
\textbf{Zhijiang Guo}$^{8, 6}$\thanks{Correspondence Authors},
\textbf{Xiao-Ming Wu}$^{1\dagger}$ \\
$^1$The Hong Kong Polytechnic University \quad $^2$Copula Lab \\
$^3$University College London \quad $^4$The Chinese University of Hong Kong, Shenzhen \\
$^5$Singapore University of Technology and Design \\
$^6$The Hong Kong University of Science and Technology \\
$^7$The Chinese University of Hong Kong \\
$^8$The Hong Kong University of Science and Technology (Guangzhou) \\
\href{mailto:zaoyu22.chen@connect.polyu.hk}{\texttt{zaoyu22.chen@connect.polyu.hk}} \quad
\href{mailto:xiao-ming.wu@polyu.edu.hk}{\texttt{xiao-ming.wu@polyu.edu.hk}}
}

\begin{document}
\maketitle

\begin{abstract}
Large language models (LLMs) can generate code from natural language, but the extent to which they capture intended program behavior remains unclear. Executable behavioral specifications, defined via preconditions and postconditions, provide a concrete means to assess such understanding. However, existing work on specification generation is limited in evaluation methodology, task settings, and specification expressiveness.
We introduce \benchname{}, a benchmark for executable behavioral specification generation under an execution-based evaluation protocol. \benchname{} supports both function-level and repository-level tasks and encodes specifications as executable Python functions. Constructed from diverse real-world codebases, it enables a realistic assessment of both correctness, by accepting valid behaviors, and completeness, by rejecting invalid behaviors.
Evaluating 15 state-of-the-art LLMs on \benchname{}, we observe a sharp performance degradation on repository-level tasks, where the best model attains only a 20.2\% pass rate. We further find that specification generation is substantially more challenging than code generation, indicating that strong coding performance does not necessarily reflect deep understanding of intended program semantics.
Our data and code are available at \url{https://github.com/SparksofAGI/CodeSpecBench}.
\end{abstract}

\section{Introduction}

Large language models (LLMs) have become integral to modern software development~\citep{10.1145/3796507} and can generate code directly from natural-language descriptions~\citep{10.1145/3747588}. \emph{Can LLMs also formalize intended program behavior as executable preconditions and postconditions?}

\begin{figure*}[t]
\centering
\newlength{\motivationpanelheight}
\setlength{\motivationpanelheight}{1.55in}
\begin{subfigure}[t]{0.56\textwidth}
\begin{minipage}[t][\motivationpanelheight][c]{0.31\linewidth}
\small
\textbf{Problem}
\vspace{1pt}

\setlength{\FrameSep}{2pt}
\begingroup
\begin{framed}
{\raggedright\scriptsize\ttfamily\color{deepred}
Given a valid (IPv4) IP address, return a defanged version. Replace every period . with [.].\par}
\end{framed}
\endgroup
\end{minipage}
\hfill
\begin{minipage}[t][\motivationpanelheight][t]{0.65\linewidth}
\small
\textbf{Correct solution}
\vspace{1pt}
\begin{Python}
class Solution:
    def defangIPaddr(self, address):
        return address.replace('.', '[.]')
\end{Python}
\vspace{3pt}
\textbf{Generated specification}
\vspace{1pt}
\begin{Python}
def postconditions(address, result):
    expected = address.replace('.', '[.]')
    assert isinstance(result, str)
    assert result == expected
    assert '.' not in result
\end{Python}
\end{minipage}
\caption{
A model can produce a correct solution while still generating an \textbf{incorrect} specification.}
\end{subfigure}
\hfill
\begin{subfigure}[t]{0.42\textwidth}
\begin{minipage}[t][\motivationpanelheight][c]{\linewidth}
\centering
\includegraphics[width=\linewidth]{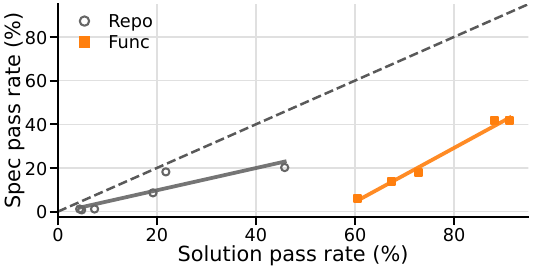}
\end{minipage}
\caption{
Strict pass rates. Lines show linear fits.}
\end{subfigure}
\caption{A valid IP-defanging solution is paired with a specification that rejects valid outputs containing \texttt{[.]} (left). Strict specification pass rates are lower than strict solution pass rates on Func tasks and Repo issues (right).}
\label{fig-motivation-example}
\end{figure*}

Executable behavioral specifications express behavioral requirements as checks that can be run against a program~\citep{10.1007/978-3-030-29026-9_22}. They characterize program behavior through preconditions and postconditions~\citep{161279}. Preconditions delimit the admissible inputs and program states for a function, and postconditions specify what must hold for the outputs and resulting program states upon termination~\citep{11029962}. In the left panel of Figure~\ref{fig-motivation-example}, a model produces a passing solution, yet its generated specification rejects a valid output. The right panel shows lower strict pass rates for specifications across the Func and Repo comparisons. At the same time, executable behavioral specifications support not only systematic evaluation that can inform and improve code generation, but also solution verification and more precise, unambiguous communication between agents.

However, the ability of LLMs to generate executable behavioral specifications remains underexplored. Existing approaches have important limitations in evaluation, task setting, and specification scope. For example, FormalBench~\citep{le-cong-etal-2025-llms} and VERINA~\citep{ye2025verina} both rely on deductive verification and exclude repository-level tasks. FormalBench depends on the OpenJML verifier, whose results can be unstable under missing invariants and complex control flow, and its specifications are derived from code rather than natural-language descriptions. VERINA relies on Lean, which is constrained by the capabilities of LLM-based theorem provers, and is limited by its relatively small dataset; moreover, its specifications target formal verification rather than capturing behavioral constraints. In contrast, nl2postcond~\citep{10.1145/3660791} uses execution-based evaluation but is restricted to relatively simple datasets, supports only limited assertion forms, and does not consider preconditions. As a result, existing approaches offer only limited evidence regarding how well LLMs can generate executable behavioral specifications in realistic settings.

To address this gap, we introduce \benchname{}, a benchmark for executable behavioral specification generation. Unlike prior work, it adopts an execution-based evaluation protocol and covers both function-level (Figure~\ref{fig-func-spec-example}) and repository-level (Figure~\ref{fig-repo-spec-example}) tasks. 
\benchname{} includes both preconditions and postconditions, expressed as executable Python functions, allowing flexible and expressive behavioral constraints. It is built from a larger and more diverse set of real-world task sources, yielding more complex and realistic scenarios than prior small-scale or simplified datasets. Correctness measures acceptance of valid behaviors, and completeness measures rejection of invalid behaviors.

We evaluate 15 state-of-the-art LLMs on \benchname{} and observe that, although models attain moderate performance on self-contained, function-level tasks, their performance degrades sharply in repository-level settings. In this more complex scenario, the best-performing model, Claude-4.5-Sonnet, achieves a 20.2\% strict pass rate (Table~\ref{tab-speccodebench-experi-results}), with errors dominated by dependency-resolution failures (Section~\ref{sec-spec-errors}). On \benchname{}-Func, the gap between solution and specification pass rates shows that expressing intended behavior as executable preconditions and postconditions poses additional challenges for these models.

\section{Related Work}
\noindent\textbf{Code Generation.}
Code generation has been studied through benchmarks at different granularities. 
\textbf{Function-level benchmarks} such as HumanEval~\citep{chen2021evaluating}, MBPP~\citep{austin2021program}, MHPP~\citep{dai2025mhppexploringcapabilitieslimitations}, LiveCodeBench~\citep{jain2024livecodebench}, EffiBench-X~\citep{qing2025effibench}, and LiveCodeBench Pro~\citep{zheng2025livecodebenchproolympiadmedalists} evaluate whether models can generate a correct function from a natural-language description. 
At a larger scale, \textbf{repository-level benchmarks} including SWE-bench~\citep{jimenez2024swebench}, EvoCodeBench~\citep{li2024evocodebench}, CrossCodeEval~\citep{ding2023crosscodeeval}, RepoBench~\citep{liu2024repobench}, and RepoCoder~\citep{zhang2023repocoder} assess generation in realistic multi-file projects, requiring reasoning over dependencies and producing patches that pass project-specific tests.
Unlike these solution-generation benchmarks, \benchname{} focuses on specification generation: models must produce explicit, executable preconditions and postconditions that capture intended behavior rather than directly implement solutions.

\begin{table*}[t!]
\centering
\small
\setlength{\tabcolsep}{3pt}
\begin{tabular}{>{\bfseries}lccccc}
\toprule
\textbf{Benchmark} & \textbf{Func/Repo} & \textbf{Pre/Post} & \textbf{Verify} & \textbf{Language} & \textbf{Repr} \\
\midrule
VERINA~\citep{ye2025verina}     & \ding{51}/\ding{55} & \ding{51}/\ding{51} & Static    & Lean    & Proposition \\
AutoSpec~\citep{10.1007/978-3-031-65630-9_16}       & \ding{51}/\ding{51} & \ding{51}/\ding{51} & Static    & ACSL     & Annotation \\
SpecGenBench~\citep{11029962}   & \ding{51}/\ding{55} & \ding{51}/\ding{51} & Static    & JML      & Annotation \\
FormalBench~\citep{le-cong-etal-2025-llms}    & \ding{51}/\ding{55} & \ding{51}/\ding{51} & Static    & JML      & Annotation \\
nl2postcond~\citep{10.1145/3660791}    & \ding{51}/\ding{51} & \ding{55}/\ding{51} & Execution & Python/Java  & Assertion \\
\midrule
\benchname{}   & \ding{51}/\ding{51} & \ding{51}/\ding{51} & Execution & Python  & Function \\
\bottomrule
\end{tabular}
\caption{Comparison of \benchname{} with representative specification-generation benchmarks along key dimensions, including granularity (function vs.\ repository), specification coverage (preconditions and postconditions), verification paradigm (static vs.\ execution-based), specification language and representation.}

\label{tab-benchmark-comparison}
\end{table*}

\noindent\textbf{Specification Generation.} Existing work on specification generation spans several directions. Classical automated methods, often based on template reuse or execution-trace mining~\citep{7371998,11029962}, typically infer specifications in predefined forms. More recent studies use large language models to generate specifications in formal languages such as Verus~\citep{shefer2025llmsenableverificationmainstream,deng2025verifythisbenchgeneratingcodespecifications}, Dafny~\citep{10918369} (e.g., DafnyBench~\citep{loughridge2025dafnybench}), Alloy~\citep{alhanahnah2025empiricalevaluationpretrainedlarge}, and Lean~\citep{thakur2025clever,ye2025verina}. Other benchmarks, including AutoSpec~\citep{10.1007/978-3-031-65630-9_16}, SpecGenBench~\citep{11029962}, and FormalBench~\citep{le-cong-etal-2025-llms}, evaluate specification generation in verification-oriented settings with deductive verifiers such as Frama-C or OpenJML. There is also growing interest in translating natural-language descriptions into assertions or postconditions, as in nl2postcond~\citep{10.1145/3660791}.

\benchname{} jointly evaluates executable Python preconditions and
postconditions on valid and invalid behaviors at both the function
and repository levels (Table~\ref{tab-benchmark-comparison}).
The minimum-swaps case shows that a postcondition can reject all
collected incorrect outputs while the precondition misses a required
input constraint (Appendix~\ref{sec-func-partial-specs}).
The cell-reachability case shows that a passing solution can be paired
with a postcondition that accepts an incorrect output
(Appendix~\ref{sec-spec-vs-solution-table}).
The Django case illustrates how a repository-level specification can
capture pre-call state and check the required state change after
execution (Appendix~\ref{sec-human-audit-repo}).

\section{Executable Behavioral Specification Generation and Evaluation}
\subsection{Specification Generation}
\label{sec-formal-specgen}
\begin{figure*}
\centering
\includegraphics[width=\textwidth]{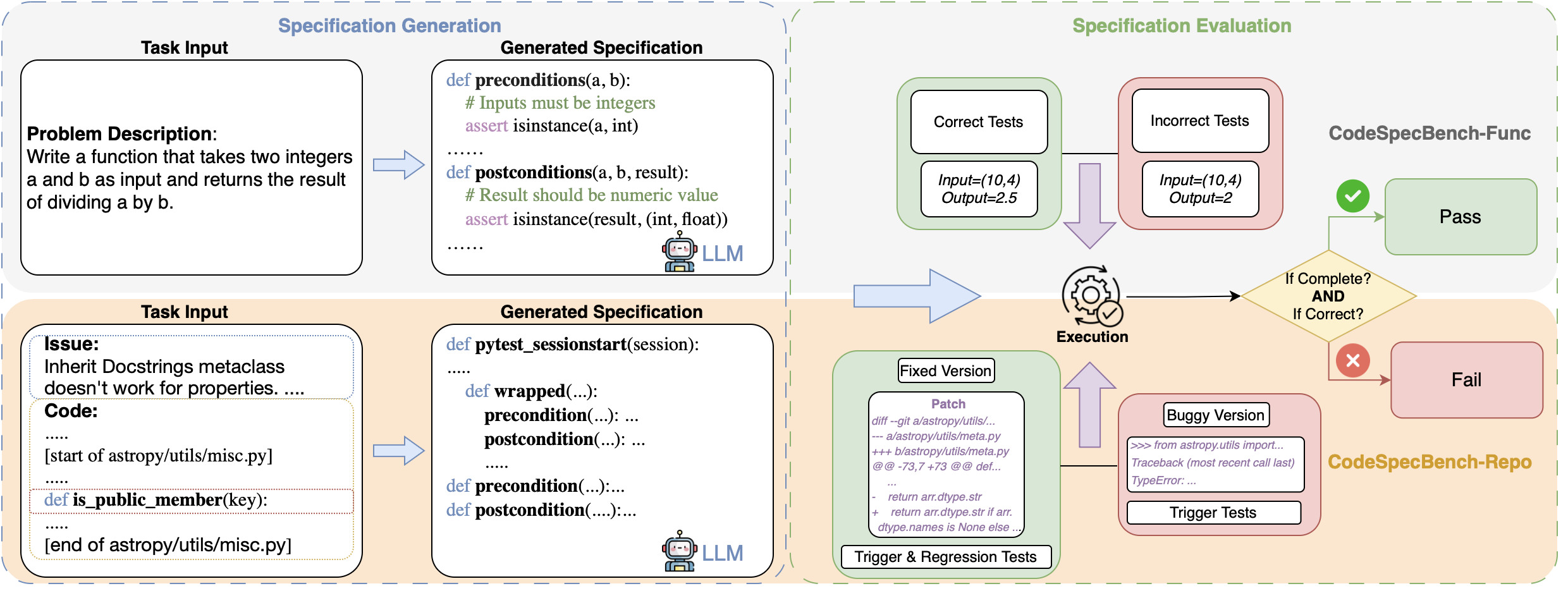}
\caption{Overview of generation and evaluation in \benchname{}.
In \benchname{}-Func (top), the specification is generated from a natural language problem description (top left) and executed on correct and incorrect test cases to evaluate correctness and completeness (top right).
In \benchname{}-Repo (bottom), the specification is generated from a natural language issue description and relevant repository code (bottom left). Correctness is evaluated using trigger and regression tests on the fixed repository, while completeness is evaluated using trigger tests on the buggy repository (bottom right).}
\label{fig-main-graph}
\end{figure*}

We study executable behavioral specification generation as the task of translating natural-language problem descriptions into executable Python specification functions. We represent these specifications in Python for three practical reasons:
(1) function-based specifications provide a flexible way to express preconditions and postconditions in a single executable form; 
(2) executable specifications support direct evaluation against curated test cases in both function-level and repository-level settings; and 
(3) Python is widely used in contemporary code benchmarks and is well supported by LLMs~\citep{twist2025studyllmspreferenceslibraries}.

In this task, an executable behavioral specification consists of a precondition and a postcondition, as shown in Figure~\ref{fig-main-graph}:

\textbf{Preconditions} specify the conditions that a function’s inputs and relevant program state must satisfy before execution, defining the domain in which the function’s behavior is well-defined.
They typically constrain input types and value ranges, and may also include assumptions about program state, such as object fields, global variables, configuration settings, or external resources (e.g., open files or network connections).

\textbf{Postconditions} specify the properties that must hold after a function executes successfully, assuming its preconditions are satisfied, thereby defining its expected behavior.
They constrain the outputs and resulting program state, often describing relationships between inputs and outputs (e.g., ordering or functional dependencies), as well as which variables or object fields may be modified.

We consider specification generation at two levels: the function level and the repository level. 
This distinction allows us to evaluate whether a model can generate executable behavioral specifications not only for self-contained programming tasks, but also for realistic codebases with state and cross-file interactions. 
The core objective is shared across the two settings, while the available context and the required form of the generated specifications differ accordingly. 
Figure~\ref{fig-main-graph} presents an overview of the specification generation workflow and highlights the distinction between these two settings.

\textbf{At the function level}, each task is a self-contained programming problem with no external dependencies. The model receives only a natural-language description of the task as input. Its required output is a pair of executable Python functions, a precondition and a postcondition, which together constitute a complete executable behavioral specification for the task. The specification checks input validity and the relationship between inputs and outputs.

\textbf{At the repository level}, each task is situated within a realistic codebase consisting of multiple interacting files. The model is provided with a natural-language description of an issue, the relevant repository code context, and the signatures of the functions that are modified by the issue. Its required output is a set of executable Python preconditions and postconditions for the issue-relevant functions. In this setting, the specifications must not only capture input--output behavior, but also constrain the relevant program state and how it is allowed to change during execution.

\subsection{Specification Evaluation}

\textbf{At the function level}, we evaluate generated specifications on the 2{,}347 retained tasks using an execution-based testing framework adapted from the HumanEval codebase~\citep{chen2021evaluating}. The specifications are executed against curated sets of correct and incorrect test cases, as described in Section~\ref{sec-dataset-construction}.

\textbf{At the repository level}, evaluation is performed by executing the model-generated specifications within the context of a real-world codebase. We use an execution-based instrumentation strategy within the SWE-bench execution framework~\citep{jimenez2024swebench}, in which the generated specifications are dynamically injected around the issue-relevant function(s). Concretely, before the execution of an issue-relevant target function, the precondition is invoked to check the runtime inputs and relevant program state. After the original function body executes, the postcondition is invoked to evaluate the inputs, outputs, and any modified state. More details are provided in Appendix~\ref{sec-repo-spec-injection}.

\textbf{Evaluation Metrics.} We compute all three strict metrics with respect to the valid and invalid tests collected for each task.

\noindent\textbullet\ \textbf{Correctness.} A specification satisfies strict correctness if it accepts every collected test labeled valid. At the function level, these tests check valid inputs and their reference outputs (Section~\ref{sec-dataset-construction}). At the repository level, they consist of trigger and regression tests executed on the fixed repository.

\noindent\textbullet\ \textbf{Completeness.} A specification satisfies strict completeness if it rejects every collected test labeled invalid. At the function level, these tests check inputs and outputs labeled invalid during dataset construction (Section~\ref{sec-dataset-construction}). At the repository level, they consist of trigger tests executed on the buggy repository.

\noindent\textbullet\ \textbf{Strict Pass Rate.} The proportion of evaluated tasks for which the generated specification satisfies both criteria.

\noindent\textbullet\ \textbf{Fractional Pass Rate.} We take the harmonic mean of the fractions of valid tests accepted and invalid tests rejected. On Func, we compute this per task and then average. On Repo, we average the two fractions across issues before taking their harmonic mean (Appendix~\ref{sec-fraction-metric}).

\section{Proposed \benchname{}}
\benchname{} consists of two evaluation levels: \benchname{}-Func, which targets function-level tasks, and \benchname{}-Repo, which focuses on repository-level issues.

\subsection{\benchname{}-Func}
\label{sec-dataset-construction}
\benchname{}-Func focuses on tasks that are self-contained and largely specified by their natural-language descriptions. It is built upon the publicly available LeetCodeDataset~\citep{xia2025leetcodedatasettemporaldatasetrobust}, 
which comprises 2{,}869 programming problems of varying difficulty levels and spans a broad range of algorithmic domains, including arrays, dynamic programming, and string processing.
From the initial 2{,}494 publicly accessible problems, we excluded 147 candidate multi-answer tasks and retained 2{,}347 problems. We screened task descriptions for arbitrary valid outputs, order-insensitive outputs, and explicitly accepted alternatives. On such tasks, an output that differs from the reference can still satisfy the problem. Appendix~\ref{sec-multianswer-sensitivity} reports the effect of this exclusion.
While many recent benchmarks, such as LiveCodeBench~\citep{jain2024livecodebench}, include on average no more than 20 test cases per problem, we leverage LLMs to construct over 200 checks per problem on average to improve reliability (see Table~\ref{tab-dataset-statistics}).

\noindent\textbf{LLM-based Test Input Generation.}
\begin{figure*}
\centering
\includegraphics[width=0.95\linewidth]{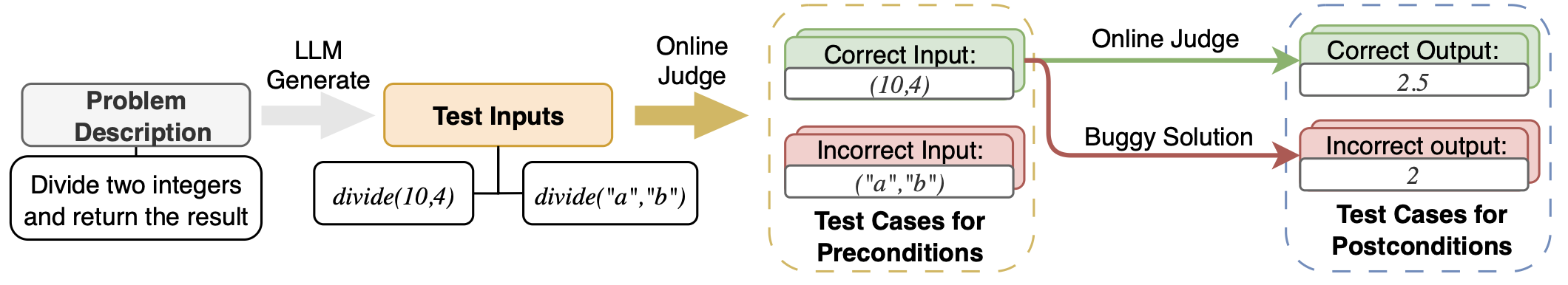}
\caption{Illustration of test case construction for \benchname{}-Func.}
\label{fig-dataset-construction}
\end{figure*}

To increase the diversity and coverage of generated test cases, we employ multiple LLMs, including GPT-4o~\citep{openai_gpt4o_2024}, DeepSeek-V3~\citep{deepseekai2025deepseekv3technicalreport}, and Qwen-Max~\citep{hui2024qwen25codertechnicalreport}. Using multiple models mitigates model-specific biases and promotes greater variability in both input distributions and edge-case coverage.
As shown in Figure~\ref{fig-dataset-construction}, the test case generation pipeline starts from a natural-language problem description, which is then fed into the model.
Each model is prompted using a carefully designed instruction template (see Appendix~\ref{sec-prompt-test-cases-gen}) to generate only the inputs of test cases, including both valid inputs that satisfy the intended problem constraints and intentionally invalid inputs that violate expected semantic properties.
After generation, we automatically parse the generated test inputs to extract structured parameters. We then apply a deduplication procedure to remove identical test cases.

\noindent\textbf{Online Judge-based Validation.}
We submit the processed test inputs to the official online judging (OJ) platform for validation.
As illustrated in Figure~\ref{fig-dataset-construction}, test inputs rejected by the OJ system (e.g., \texttt{("a","b")}) are labeled as incorrect test inputs. These inputs are used as incorrect test cases to evaluate precondition completeness.
The OJ accepts valid inputs such as \texttt{(10, 4)} and returns a reference output such as \texttt{2.5}. We use the accepted input to test precondition correctness and its input--output pair to test postcondition correctness.
We repeat input generation and validation to collect up to 50 valid and 50 invalid inputs per problem. Some problems have small valid input domains, for example $n \in \{0,1,\dots,7\}$.
We categorized the inputs rejected by the online judge using rules derived from the corresponding task descriptions.
Among the rejected inputs analyzed, 39.2\% had type or format errors, 41.0\% contained values outside the allowed ranges, 16.4\% violated size or shape constraints, and 3.4\% violated constraints relating multiple inputs.

\noindent\textbf{Construction of Incorrect Output.}
We construct incorrect output pairs by sampling buggy solutions from LLMs. Specifically, we employ three models with different capability levels, Qwen3-8B, Qwen3-32B, and GPT-5-mini, to generate candidate solutions. For each problem, we prompt the models to produce 10 solutions (see Appendix~\ref{sec-prompt-solution-gen} for the prompt template).
We validate each solution using the valid input--output pairs obtained from the OJ. For each input, we execute the solution and collect outputs that differ from the reference as negative postcondition checks. This rule can mislabel valid alternatives. The description-based exclusion above reduces this ambiguity. The retained tasks have an average of 67.3 incorrect-output checks per problem.
To ensure the quality of the generated test cases, we further conduct a quantitative evaluation by measuring statement and branch coverage, with average statement coverage of 96.4\% and branch coverage of 93.8\% across the 2{,}346 retained tasks with coverage reports.
Additional quality assurance details are provided in Appendix~\ref{sec-quality-assurance}.

\subsection{\benchname{}-Repo}
\benchname{}-Repo requires models to understand repository structure, module dependencies, and broader execution behavior when generating specifications.
For repository-level evaluation, we adopt SWE-bench Verified~\citep{jimenez2024swebench}, which provides challenging real-world software issues grounded in large, multi-file Python repositories.
SWE-bench Verified contains \textbf{500} curated issues across \textbf{12} widely used open-source projects (e.g., \texttt{astropy}, \texttt{matplotlib}, \texttt{sympy}), with complex intra-project dependencies.
Each instance consists of a natural-language bug report or feature request, the relevant code context from the associated repository, the buggy (pre-patch) and fixed (post-patch) versions of the repository, and trigger tests and regression tests, which correspond to the FAIL\_TO\_PASS and PASS\_TO\_PASS tests defined in SWE-bench.

Prior work~\citep{10.5555/3780338.3780368,10.5555/3737916.3740517} has pointed out that SWE-bench suffers from insufficient test coverage. To address this issue, UTBoost~\citep{yu-etal-2025-utboost} identifies instances with insufficient test cases and augments them using LLM-generated tests to enable more rigorous and accurate verification. We adopt the augmented test cases from this work for evaluation.
\begin{table}[tb]
\centering
\begin{minipage}[t]{0.45\textwidth}
\vspace{0pt}
    \centering
    \small
\setlength{\tabcolsep}{3pt}
\begin{tabular}{>{\bfseries}lcc}
\toprule
\textbf{Metric} & \textbf{Func} & \textbf{Repo} \\
\midrule
Num of tasks & 2{,}347 & 500 \\
\midrule
Avg checks per task & 215.5 & 123.3 \\
\quad Correct checks & 98.3 & -- \\
\quad Incorrect checks & 117.1 & -- \\
\quad Trigger tests & -- & 3.1 \\
\quad Regression tests & -- & 120.3 \\
\midrule
Avg prompt tokens & 511.7 & 19.7k \\
Avg statement coverage & 96.4\% & -- \\
Avg branch coverage & 93.8\% & -- \\
\bottomrule
\end{tabular}
\caption{Statistics for \benchname{}-Func and \benchname{}-Repo.}
\label{tab-dataset-statistics} 
\end{minipage}
\hfill %
\begin{minipage}[t]{0.5\textwidth}
\vspace{0pt}
\centering
\small
\setlength{\tabcolsep}{3pt}
\begin{threeparttable}
\begin{tabular}{lccc}
\toprule
\textbf{Model} & \textbf{OS} & \textbf{RM} & \textbf{CL} \\
\midrule
\textbf{Qwen3}$^{*}$ & \ding{51}  & \ding{51} & 32k / 128k \\
\textbf{QWQ-32B} & \ding{51} & \ding{51} & 32k \\
\textbf{DeepSeek-V3.2} & \ding{51} & \ding{51} & 128k \\
\textbf{GPT-OSS} & \ding{51} & \ding{51} & 128k \\
\textbf{Claude-4.5-Sonnet} & \ding{55} & \ding{51} & 200k \\
\textbf{GPT-5-mini} & \ding{55} & \ding{51} & 272k \\
\textbf{GPT-5} & \ding{55} & \ding{51} & 272k \\
\textbf{Gemini-2.5-Pro} & \ding{55} & \ding{51} & 128k \\
\textbf{Gemini-2.5-Flash} & \ding{55} & \ding{51} & 128k \\
\bottomrule
\end{tabular}
\begin{tablenotes}
\footnotesize
\item[*] 32k for models $\leq$4B; 128k for $\geq$8B.
\end{tablenotes}
\end{threeparttable}
\caption{Selected LLMs. OS denotes open-source models, RM denotes reasoning models, and CL gives context length in tokens.}
\label{tab_llms}
\end{minipage}
\end{table}

\section{Experiments and Results}
\subsection{Experimental Setup}
We evaluate a diverse set of contemporary large language models on \benchname{}, as summarized in Table~\ref{tab_llms}. Deployment details are provided in Appendix~\ref{sec-model-deployment-details}. Prompt used for generating executable behavioral specification is provided in Appendix~\ref{sec-prompt-specification-gen}.

\subsection{Main Results}
\definecolor{scbf}{RGB}{235,242,255}  %
\definecolor{scbr}{RGB}{255,239,230}  %

\definecolor{groupPink}{RGB}{255,230,230}
\definecolor{groupGreen}{RGB}{220,255,220}

\begin{table*}[tb]
\centering
\small
\setlength{\tabcolsep}{3.5pt}
\begin{tabular}{l c c c c c c c c c}

\toprule
\multirow{2}{*}{\textbf{Model}} & \multirow{2}{*}{\textbf{Reasoning}} 
& \multicolumn{4}{>{\columncolor{scbf}}c}{\textbf{\benchname{}-Func}}
& \multicolumn{4}{>{\columncolor{scbr}}c}{\textbf{\benchname{}-Repo}} \\
\cmidrule(lr){3-6} \cmidrule(lr){7-10}
 &  
 & \cellcolor{scbf}\textbf{Correct} 
 & \cellcolor{scbf}\textbf{Complete} 
 & \cellcolor{scbf}\textbf{Strict}
 & \cellcolor{scbf}\textbf{Fractional}
 & \cellcolor{scbr}\textbf{Correct} 
 & \cellcolor{scbr}\textbf{Complete} 
 & \cellcolor{scbr}\textbf{Strict}
 & \cellcolor{scbr}\textbf{Fractional} \\
\midrule

\rowcolor{groupGreen}
\multicolumn{10}{c}{\textit{Open-Weight Models}} \\

\multirow{2}{*}{\textbf{Qwen3-0.6B}} & \ding{55} & 8.4 & 11.7 & 0.2 & 7.0 & 0.0 & 0.0 & 0.0 & 0.0 \\
          & \ding{51} & 5.1 & 17.3 & 0.3 & 4.0 & 0.0 & 0.0 & 0.0 & 0.0 \\
          
\multirow{2}{*}{\textbf{Qwen3-1.7B}} & \ding{55} & 8.3 & 20.9 & 0.8 & 10.6 & 0.2 & 0.2 & 0.0 & 0.3 \\
           & \ding{51} & 15.0 & 11.2 & 1.3 & 16.8 & 0.0 & 0.0 & 0.0 & 0.0 \\
           
\multirow{2}{*}{\textbf{Qwen3-4B}} & \ding{55} & 44.5 & 20.0 & 5.7 & 45.5 & 0.2 & 1.2 & 0.0 & 0.9 \\
         & \ding{51} & 71.8 & 22.5 & 16.0 & 78.5 & 0.6 & 1.2 & 0.2 & 1.4 \\
         
\multirow{2}{*}{\textbf{Qwen3-8B}} & \ding{55} & 48.8 & 17.3 & 6.0 & 51.1 & 1.6 & 10.0 & 0.8 & 8.8 \\
         & \ding{51} & 75.7 & 26.3 & 17.3 & 74.0 & 2.0 & 7.2 & 0.0 & 7.0 \\
         
\multirow{2}{*}{\textbf{Qwen3-14B}} & \ding{55} & 63.4 & 24.5 & 13.9 & 66.1 & 4.8 & 17.4 & 1.2 & 17.0 \\
          & \ding{51} & 80.1 & 35.3 & 27.3 & \underline{84.5} & 6.0 & 17.8 & 1.2 & 18.8 \\
          
\multirow{2}{*}{\textbf{Qwen3-32B}} & \ding{55} & 65.8 & 28.6 & 18.0 & 70.5 & 3.2 & 8.4 & 1.2 & 8.7 \\
          & \ding{51} & 82.0 & 35.6 & 26.6 & 82.6 & 8.8 & 15.6 & 2.0 & 18.2 \\
          
\textbf{QWQ-32B} & \ding{51} & \textbf{83.3} & 38.2 & 29.7 & 83.1 & \underline{10.6} & 25.0 & \underline{3.6} & 25.6 \\
\textbf{GPT-OSS-20B} & \ding{51} & 73.5 & \underline{53.0} & \underline{39.0} & 81.2 & 9.4 & \underline{31.6} & 3.4 & 34.5 \\
\textbf{GPT-OSS-120B} & \ding{51} & \underline{82.5} & \textbf{53.4} & \textbf{44.0} & \textbf{88.5} & 8.6 & \textbf{36.2} & 3.2 & \textbf{35.9} \\
\textbf{DeepSeek-V3.2} & \ding{55} & 72.8 & 35.4 & 24.9 & 76.4 & \textbf{20.2} & 30.2 & \textbf{6.8} & \underline{35.4} \\
\midrule

\rowcolor{groupPink}
\multicolumn{10}{c}{\textit{Proprietary Models}} \\

\textbf{Gemini-2.5-Flash} & \ding{51} & 83.4 & 45.3 & 37.8 & 87.5 & 9.2 & 17.6 & 2.8 & 21.5 \\
\textbf{Gemini-2.5-Pro} & \ding{51} & \textbf{88.0} & \underline{54.6} & \underline{47.8} & \textbf{92.6} & \underline{30.8} & 53.0 & \underline{18.2} & \underline{55.9} \\
\textbf{GPT-5-mini} & \ding{51} & 81.9 & \textbf{61.6} & \textbf{48.6} & 86.8 & 19.6 & 45.8 & 9.6 & 45.7 \\
\textbf{GPT-5} & \ding{51} & 74.9 & 54.0 & 41.6 & 87.5 & 18.8 & \textbf{58.6} & 8.6 & 55.5 \\
\textbf{Claude-4.5-Sonnet} & \ding{51} & \underline{84.8} & 47.5 & 41.8 & \underline{91.3} & \textbf{37.4} & \underline{57.2} & \textbf{20.2} & \textbf{62.5} \\
\bottomrule
\end{tabular}
\caption{Specification generation results on Func and Repo (percentages). Correct and Complete report strict correctness and completeness. Strict and Fractional denote the two pass rates. Reasoning indicates whether thinking mode is enabled.}
\label{tab-speccodebench-experi-results}
\end{table*}

\paragraph{How well do LLMs generate executable behavioral specifications?}
\noindent\emph{Pass rate is moderate on \benchname{}-Func but drops sharply on \benchname{}-Repo.}
As shown in Table~\ref{tab-speccodebench-experi-results}, the best strict pass rates on \benchname{}-Func are 48.6\% (GPT-5-mini), 47.8\% (Gemini-2.5-Pro), and 44.0\% (GPT-OSS-120B). Performance drops sharply on \benchname{}-Repo, where the best results are 20.2\% (Claude-4.5-Sonnet), 18.2\% (Gemini-2.5-Pro), and 9.6\% (GPT-5-mini). Open-weight models score lower, with Qwen3-32B reaching 2.0\% and DeepSeek-V3.2 reaching 6.8\%.

Fractional pass captures partial success on the collected tests. On Repo, Claude-4.5-Sonnet reaches 62.5\% fractional pass. GPT-5-mini scores higher than GPT-5 on strict pass (9.6\% vs.\ 8.6\%). The ordering reverses on fractional pass, where GPT-5 scores 55.5\% and GPT-5-mini scores 45.7\%. Appendix~\ref{sec-bootstrap-ci} reports task-level bootstrap 95\% confidence intervals and repeated runs. Across five runs, Qwen3-32B and GPT-5-mini consistently had lower strict pass rates on Repo, and GPT-5-mini varied more across runs on Repo than on Func.

\noindent\emph{Scaling and reasoning help on \benchname{}-Func, while proprietary models dominate \benchname{}-Repo.}
Within Qwen3, \benchname{}-Func pass rate rises from 0.2\% (0.6B without reasoning) to 18.0\% (32B without reasoning) and 26.6\% (32B with reasoning). Reasoning substantially helps mid-to-large models, e.g., Qwen3-8B improves from 6.0\% to 17.3\% and Qwen3-14B from 13.9\% to 27.3\%, but brings little benefit to smaller models. On \benchname{}-Repo, these gains are weak: Qwen3-32B peaks at 2.0\%, while proprietary models lead with Claude-4.5-Sonnet (20.2\%) and Gemini-2.5-Pro (18.2\%).

\paragraph{How do generated specifications trade off correctness and completeness?}
\noindent\emph{On \benchname{}-Func, correctness usually exceeds completeness, indicating a tendency toward overly loose specifications.}
Table~\ref{tab-speccodebench-experi-results} shows a consistent and substantial gap between correctness and completeness across most models. For instance, QWQ-32B achieves 83.3\% correctness and 38.2\% completeness, and Qwen3-32B (with reasoning) shows a similar disparity (82.0\% vs.\ 35.6\%). Even strong proprietary models such as Gemini-2.5-Pro (88.0\% vs.\ 54.6\%) and GPT-5-mini (81.9\% vs.\ 61.6\%) exhibit the same pattern.
Notably, the pass rate is much closer to completeness than to correctness (e.g., Gemini-2.5-Pro achieves 47.8\% pass, 54.6\% completeness, and 88.0\% correctness, while GPT-5-mini achieves 48.6\%, 61.6\%, and 81.9\%, respectively). This indicates that on self-contained, function-level tasks, models tend to produce permissive specifications that admit many valid behaviors but fail to rule out enough invalid ones.

\noindent\emph{On \benchname{}-Repo, strict completeness generally exceeds strict correctness.}
For example, Gemini-2.5-Pro achieves 53.0\% completeness and 30.8\% correctness, while GPT-5 attains 58.6\% completeness and 18.8\% correctness (Table~\ref{tab-speccodebench-experi-results}). Fewer issues satisfy every valid-behavior check on the fixed repository than satisfy every invalid-behavior check on the buggy repository. Failures on the fixed repository include overly restrictive conditions and instrumentation failures.
A manual audit further confirms that most passed specifications are meaningful and tailored to the corresponding problem or issue, while failed cases often involve weak checks, overly restrictive assumptions, or failures in resolving the target hook (Appendix~\ref{sec-human-audit-specs}).

\subsection{Analysis and Discussion}
\paragraph{Specification failure modes.}
\label{sec-spec-errors}
We conduct a fine-grained error analysis (Appendix~\ref{sec-error-taxonomy}), which suggests that errors primarily arise from difficulty in specifying input constraints and grounding them in code context. For the two models in the function-level error analysis, \emph{Type} and \emph{Range Violations} account for 48--54\% of classified failures (Table~\ref{tab-error-distribution-func}), indicating mis-specified input domains. On \benchname{}-Repo, integration failures include unresolved dependencies and hooks that never run. Specification errors include overly restrictive conditions that reject valid behaviors. Among Claude-4.5-Sonnet's classified failures, dependency resolution accounts for 33.4\%, untriggered hooks for 12.4\%, and overly restrictive specifications for 18.7\% (Table~\ref{tab-error-distribution-repo}).

\begin{table}[tb]
\centering
\small
\begin{tabular}{lrr}
\toprule
\textbf{Model} & \textbf{End-to-end} & \textbf{Scaffold given} \\
\midrule
\textbf{Gemini-2.5-Pro} & 45.5 & 53.0 \\
\textbf{GPT-5-mini} & 24.2 & 59.6 \\
\textbf{GPT-5} & 21.7 & 61.1 \\
\textbf{Claude-4.5-Sonnet} & 50.5 & 62.6 \\
\bottomrule
\end{tabular}
\caption{Repo strict pass rates (\%) with and without a provided scaffold.}
\label{tab-repo-scaffold-ablation}
\end{table}

\paragraph{Repository scaffold ablation.}
Models must install working repository hooks and write executable preconditions and postconditions on \benchname{}-Repo. On 198 issues where we could extract a scaffold from a passing specification, we compare end-to-end generation with generation of only the condition bodies given the scaffold. GPT-5-mini's strict pass rate rises from 24.2\% to 59.6\% (Table~\ref{tab-repo-scaffold-ablation}). With the scaffold provided, strict pass ranges from 53.0\% to 62.6\% across the four models, indicating that writing the condition bodies remains challenging on this subset (Appendix~\ref{sec-repo-scaffold-ablation}).

\begin{table}[tb]
\centering
\small
\setlength{\tabcolsep}{3pt}
\begin{tabular}{>{\bfseries}lcccc}

\toprule
\textbf{Method} & \textbf{Pass@1} & \textbf{Pass@2} & \textbf{Pass@3} & \textbf{Pass@5} \\
\midrule
 & \multicolumn{4}{c}{\textit{Qwen3-8B (Non-Thinking)}} \\
 \addlinespace[2pt]
Baseline & 59.4 & 66.2 & 69.6 & 73.3 \\
Spec-guided & 62.2 & 68.2 & 70.7 & 73.9 \\

 & \multicolumn{4}{c}{\textit{Qwen3-32B (Non-Thinking)}} \\
 \addlinespace[2pt]
Baseline & 71.2 & 78.4 & 81.3 & 84.1 \\
Spec-guided & 73.9 & 79.8 & 82.2 & 85.0 \\

\bottomrule
\end{tabular}
\caption{Specification-guided re-ranking of function-level solutions (values in \%).}
\label{tab-spec-rerank-solution}
\end{table}

\paragraph{Specification-guided re-ranking.}
Table~\ref{tab-spec-rerank-solution} reports the re-ranking experiment.
Specification-guided re-ranking consistently improves performance for both models. For Qwen3-8B, $\text{pass@}1$ increases from 59.4\% to 62.2\%, while Qwen3-32B improves from 71.2\% to 73.9\%. Gains for larger $k$ are smaller but still consistent. The larger improvement on $\text{pass@}1$ suggests that generated specifications help better identify the best solution among candidates.
The experimental design is detailed in Appendix~\ref{sec-spec-reranking-design}.

\paragraph{Specification and solution pass rates.}
\begin{figure}[tb]
    \centering
    \includegraphics[width=\linewidth]{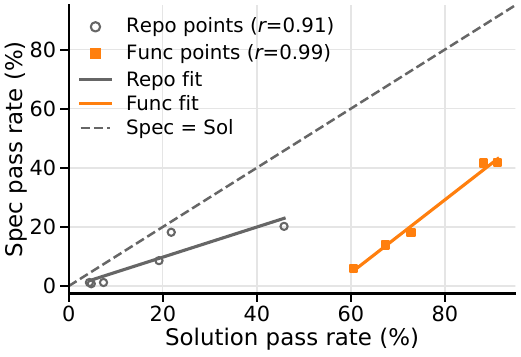}
    \caption{Strict specification and solution pass rates on Func and Repo.}
    \label{fig-spec-vs-solution}
\end{figure}

On the same 2{,}347 \benchname{}-Func tasks, all five paired models have lower strict specification pass rates than strict solution pass rates (Figure~\ref{fig-spec-vs-solution}). Across these five models, the strict rates have Pearson $r=0.99$. Fractional pass is also lower for specifications in all five pairs, with smaller gaps. For example, Claude-4.5-Sonnet achieves 41.8\% strict specification pass and 91.3\% strict solution pass, compared with 91.3\% and 95.4\% under the respective fractional metrics.

On the 500 \benchname{}-Repo issues, the patches generated in our runs have higher strict pass rates than specifications for all six paired models. For example, Claude-4.5-Sonnet passes 20.2\% of issues with specifications and resolves 45.8\% with patches. Fractional pass is higher for solutions in three of six comparisons. Appendix~\ref{sec-spec-vs-solution-table} reports the full results.

\begin{figure}[tb]
\centering
\begin{minipage}[t]{0.43\linewidth}
\footnotesize\textbf{Passing patch}\par
\begin{Python}
def __reversed__(self):
    return reversed(
        list(self.dict)
    )
\end{Python}
\end{minipage}
\hfill
\begin{minipage}[t]{0.53\linewidth}
\footnotesize\textbf{Failing precondition}\par
\begin{Python}
def precondition():
    s = OrderedSet([1, 2, 3])
    try:
        reversed(s)
    except TypeError:
        pass
    else:
        assert False
\end{Python}
\end{minipage}
\caption{Django issue 14089. The precondition rejects a successful \texttt{reversed()} call.}
\label{fig-repo-case-study}
\end{figure}

\paragraph{Repository case study.}
Figure~\ref{fig-repo-case-study} shows the generated patch and precondition for Django issue 14089, which requires \texttt{OrderedSet} to support \texttt{reversed()}. Given an ordered set containing \texttt{[1, 2, 3]}, the call should yield \texttt{3, 2, 1}. In Qwen3-32B's default solution run, the generated patch adds \texttt{\_\_reversed\_\_} and passes the SWE-bench tests. In the thinking-mode specification run, the generated precondition expects the call to raise \texttt{TypeError}, and the generated hook installs \texttt{\_\_reversed\_\_} itself. This precondition rejects the successful call on the fixed repository. A direct behavioral check would compare the returned sequence with the expected reverse order.

\section{Conclusion}

We introduced \benchname{}, a benchmark for executable behavioral specification generation with execution-based evaluation. It covers both function- and repository-level tasks, using executable Python preconditions and postconditions to assess correctness and completeness. Experiments on 15 LLMs show that this task remains challenging, especially at the repository level. In both settings, models pass the strict evaluation more often with generated solutions than with executable specifications. We hope \benchname{} will support future evaluation of this capability and encourage models that better capture program semantics.

\section*{Limitations}
Strict pass measures acceptance and rejection over the collected tests. Finite tests can miss behaviors that a specification handles incorrectly. The description-based screen removes 147 candidate multi-answer tasks, and other ambiguous cases may remain.
\benchname{} focuses on Python tasks at two granularities: self-contained function-level problems and repository-level issues from real-world open-source projects. This design covers both algorithmic reasoning and realistic project context, but software development spans a broader range of languages, ecosystems, and application domains. Future work can extend the same benchmark construction and evaluation protocol to additional programming languages, larger project types, and more specialized domains, thereby broadening the empirical picture of executable behavioral specification generation.

\section*{Ethical considerations}
Our benchmark is constructed solely from publicly available datasets and open-source repositories, with all executions conducted in controlled environments to mitigate security risks. The task involves executable behavioral specification generation for software and does not include human subjects or sensitive personal data, thereby minimizing privacy and safety concerns.

\section*{Acknowledgments}
We would like to thank Wenxuan Shi, Nianzu Zheng, Kexin Xie and Peng Tan for their support, discussions, and constructive feedback during the development of this work. This work was partially supported by the Hong Kong General Research Fund (15204225), and Project P0056021 (Parent Project P0050643) of the Otto Poon Charitable Foundation Smart Cities Research Institute, The Hong Kong Polytechnic University.

\bibliography{latex/2BibTex}

@article{chen2021evaluating,
  title={Evaluating large language models trained on code},
  author={Chen, Mark and Tworek, Jerry and Jun, Heewoo and Yuan, Qiming and Pinto, Henrique Ponde de Oliveira and Kaplan, Jared and Edwards, Harri and Burda, Yuri and Joseph, Nicholas and Brockman, Greg and others},
  journal={arXiv preprint arXiv:2107.03374},
  year={2021}
}

@inproceedings{zhang2023repocoder,
  title={RepoCoder: Repository-Level Code Completion Through Iterative Retrieval and Generation},
  author={Zhang, Fengji and Chen, Bei and Zhang, Yue and Keung, Jacky and Liu, Jin and Zan, Daoguang and Mao, Yi and Lou, Jian-Guang and Chen, Weizhu},
  booktitle={Proceedings of the 2023 Conference on Empirical Methods in Natural Language Processing},
  pages={2471--2484},
  year={2023}
}

@article{qing2025effibench,
  title={Effibench-x: A multi-language benchmark for measuring efficiency of llm-generated code},
  author={Qing, Yuhao and Zhu, Boyu and Du, Mingzhe and Guo, Zhijiang and Zhuo, Terry Yue and Zhang, Qianru and Zhang, Jie M and Cui, Heming and Yiu, Siu-Ming and Huang, Dong and others},
  journal={arXiv preprint arXiv:2505.13004},
  year={2025}
}

@article{austin2021program,
  title={Program synthesis with large language models},
  author={Austin, Jacob and Odena, Augustus and Nye, Maxwell and Bosma, Maarten and Michalewski, Henryk and Dohan, David and Jiang, Ellen and Cai, Carrie and Terry, Michael and Le, Quoc and others},
  journal={arXiv preprint arXiv:2108.07732},
  year={2021}
}

@misc{claude45_sonnet,
  title = {Claude 4.5 Sonnet},
  author = {Anthropic},
  year = {2024},
  howpublished = {\url{https://www.anthropic.com/claude/sonnet}},
  note = {Accessed: 2026-01-05}
}

@misc{openai_gpt4o_2024,
  author = {{OpenAI}},
  title = {{GPT-4o}},
  howpublished = {\url{https://openai.com/index/hello-gpt-4o}},
  year = {2024},
  note = {Accessed: 2025-02-25}
}

@misc{openai_gpt5,
  title = {Introducing GPT-5},
  author = {{OpenAI}},
  year = {2025},
  howpublished = {\url{https://openai.com/index/introducing-gpt-5/}},
  note = {Accessed: 2026-01-05}
}

@misc{google_gemini_2.5_2025,
  title = {Gemini 2.5: Our most intelligent models are getting even better},
  author = {{Google DeepMind}},
  year = {2025},
  month = {March},
  howpublished = {\url{https://blog.google/technology/google-deepmind/gemini-model-thinking-updates-march-2025/}},
  note = {Accessed: 2026-01-05}
}

@misc{hui2024qwen25codertechnicalreport,
      title={Qwen2.5-Coder Technical Report}, 
      author={Binyuan Hui and Jian Yang and Zeyu Cui and others},
      year={2024},
      eprint={2409.12186},
      archivePrefix={arXiv},
      primaryClass={cs.CL},
      url={https://arxiv.org/abs/2409.12186}, 
}

@misc{deepseekai2025deepseekv3technicalreport,
      title={DeepSeek-V3 Technical Report}, 
      author={DeepSeek-AI and Aixin Liu and Bei Feng and others},
      year={2025},
      eprint={2412.19437},
      archivePrefix={arXiv},
      primaryClass={cs.CL},
      url={https://arxiv.org/abs/2412.19437}, 
}

@article{jain2024livecodebench,
    title={LiveCodeBench: Holistic and Contamination Free Evaluation of Large Language Models for Code},
    author={Jain, Naman and Han, King and Gu, Alex and Li, Wen-Ding and Yan, Fanjia and Zhang, Tianjun and Wang, Sida and Solar-Lezama, Armando and Sen, Koushik and Stoica, Ion},
    journal={arXiv preprint arXiv:2403.07974},
    year={2024}
}

@misc{xia2025leetcodedatasettemporaldatasetrobust,
      title={LeetCodeDataset: A Temporal Dataset for Robust Evaluation and Efficient Training of Code LLMs}, 
      author={Yunhui Xia and Wei Shen and Yan Wang and Jason Klein Liu and Huifeng Sun and Siyue Wu and Jian Hu and Xiaolong Xu},
      year={2025},
      eprint={2504.14655},
      archivePrefix={arXiv},
      primaryClass={cs.LG},
      url={https://arxiv.org/abs/2504.14655}, 
}

@inproceedings{
    jimenez2024swebench,
    title={{SWE}-bench: Can Language Models Resolve Real-world Github Issues?},
    author={Carlos E Jimenez and John Yang and Alexander Wettig and Shunyu Yao and Kexin Pei and Ofir Press and Karthik R Narasimhan},
    booktitle={The Twelfth International Conference on Learning Representations},
    year={2024},
    url={https://openreview.net/forum?id=VTF8yNQM66}
}

@inproceedings{
li2024evocodebench,
title={EvoCodeBench: An Evolving Code Generation Benchmark with Domain-Specific Evaluations},
author={Jia Li and Ge Li and Xuanming Zhang and Yunfei Zhao and Yihong Dong and Zhi Jin and Binhua Li and Fei Huang and Yongbin Li},
booktitle={The Thirty-eight Conference on Neural Information Processing Systems Datasets and Benchmarks Track},
year={2024},
url={https://openreview.net/forum?id=kvjbFVHpny}
}

@inproceedings{
ding2023crosscodeeval,
title={CrossCodeEval: A Diverse and Multilingual Benchmark for Cross-File Code Completion},
author={Yangruibo Ding and Zijian Wang and Wasi Uddin Ahmad and Hantian Ding and Ming Tan and Nihal Jain and Murali Krishna Ramanathan and Ramesh Nallapati and Parminder Bhatia and Dan Roth and Bing Xiang},
booktitle={Thirty-seventh Conference on Neural Information Processing Systems Datasets and Benchmarks Track},
year={2023},
url={https://openreview.net/forum?id=wgDcbBMSfh}
}

@inproceedings{
liu2024repobench,
title={RepoBench: Benchmarking Repository-Level Code Auto-Completion Systems},
author={Tianyang Liu and Canwen Xu and Julian McAuley},
booktitle={The Twelfth International Conference on Learning Representations},
year={2024},
url={https://openreview.net/forum?id=pPjZIOuQuF}
}

@article{10.1145/3660791,
author = {Endres, Madeline and Fakhoury, Sarah and Chakraborty, Saikat and Lahiri, Shuvendu K.},
title = {Can Large Language Models Transform Natural Language Intent into Formal Method Postconditions?},
year = {2024},
issue_date = {July 2024},
publisher = {Association for Computing Machinery},
address = {New York, NY, USA},
volume = {1},
number = {FSE},
url = {https://doi.org/10.1145/3660791},
doi = {10.1145/3660791},
journal = {Proc. ACM Softw. Eng.},
month = jul,
articleno = {84},
numpages = {24}
}

@INPROCEEDINGS{11029962,
  author={Ma, Lezhi and Liu, Shangqing and Li, Yi and Xie, Xiaofei and Bu, Lei},
  booktitle={2025 IEEE/ACM 47th International Conference on Software Engineering (ICSE)}, 
  title={SpecGen: Automated Generation of Formal Program Specifications via Large Language Models}, 
  year={2025},
  volume={},
  number={},
  pages={16-28},
  doi={10.1109/ICSE55347.2025.00129}}

@inproceedings{
thakur2025clever,
title={{CLEVER}: A Curated Benchmark for Formally Verified Code Generation},
author={Amitayush Thakur and Jasper Lee and George Tsoukalas and Meghana Sistla and Matthew Zhao and Stefan Zetzsche and Greg Durrett and Yisong Yue and Swarat Chaudhuri},
booktitle={2nd AI for Math Workshop @ ICML 2025},
year={2025},
url={https://openreview.net/forum?id=pqNFDA2TFm}
}

@article{
loughridge2025dafnybench,
title={DafnyBench: A Benchmark for Formal Software Verification},
author={Chloe R Loughridge and Qinyi Sun and Seth Ahrenbach and Federico Cassano and Chuyue Sun and Ying Sheng and Anish Mudide and Md Rakib Hossain Misu and Nada Amin and Max Tegmark},
journal={Transactions on Machine Learning Research},
issn={2835-8856},
year={2025},
url={https://openreview.net/forum?id=yBgTVWccIx},
note={}
}

@inproceedings{
ye2025verina,
title={Verina: Benchmarking Verifiable Code Generation},
author={Zhe Ye and Zhengxu Yan and Jingxuan He and Timothe Kasriel and Kaiyu Yang and Dawn Song},
booktitle={2nd AI for Math Workshop @ ICML 2025},
year={2025},
url={https://openreview.net/forum?id=47601CQFri}
}

@misc{twist2025studyllmspreferenceslibraries,
      title={A Study of LLMs' Preferences for Libraries and Programming Languages}, 
      author={Lukas Twist and Jie M. Zhang and Mark Harman and Don Syme and Joost Noppen and Helen Yannakoudakis and Detlef Nauck},
      year={2025},
      eprint={2503.17181},
      archivePrefix={arXiv},
      primaryClass={cs.SE},
      url={https://arxiv.org/abs/2503.17181}, 
}

@misc{dai2025mhppexploringcapabilitieslimitations,
      title={MHPP: Exploring the Capabilities and Limitations of Language Models Beyond Basic Code Generation}, 
      author={Jianbo Dai and Jianqiao Lu and Yunlong Feng and Guangtao Zeng and Rongju Ruan and Ming Cheng and Dong Huang and Haochen Tan and Zhijiang Guo},
      year={2025},
      eprint={2405.11430},
      archivePrefix={arXiv},
      primaryClass={cs.CL},
      url={https://arxiv.org/abs/2405.11430}, 
}

@INPROCEEDINGS{7371998,
  author={Lemieux, Caroline and Park, Dennis and Beschastnikh, Ivan},
  booktitle={2015 30th IEEE/ACM International Conference on Automated Software Engineering (ASE)}, 
  title={General LTL Specification Mining (T)}, 
  year={2015},
  volume={},
  number={},
  pages={81-92},
  doi={10.1109/ASE.2015.71}}

@misc{shefer2025llmsenableverificationmainstream,
      title={Can LLMs Enable Verification in Mainstream Programming?}, 
      author={Aleksandr Shefer and Igor Engel and Stanislav Alekseev and Daniil Berezun and Ekaterina Verbitskaia and Anton Podkopaev},
      year={2025},
      eprint={2503.14183},
      archivePrefix={arXiv},
      primaryClass={cs.SE},
      url={https://arxiv.org/abs/2503.14183}, 
}

@misc{deng2025verifythisbenchgeneratingcodespecifications,
      title={VerifyThisBench: Generating Code, Specifications, and Proofs All at Once}, 
      author={Xun Deng and Sicheng Zhong and Barış Bayazıt and Andreas Veneris and Fan Long and Xujie Si},
      year={2025},
      eprint={2505.19271},
      archivePrefix={arXiv},
      primaryClass={cs.SE},
      url={https://arxiv.org/abs/2505.19271}, 
}

@INPROCEEDINGS{10918369,
  author={Lahirie, Shuvendu K.},
  booktitle={2024 Formal Methods in Computer-Aided Design (FMCAD)}, 
  title={Evaluating LLM-driven User-Intent Formalization for Verification-Aware Languages}, 
  year={2024},
  volume={},
  number={},
  pages={142-147},
  doi={10.34727/2024/isbn.978-3-85448-065-5_19}}

@inproceedings{le-cong-etal-2025-llms,
    title = "Can {LLM}s Reason About Program Semantics? A Comprehensive Evaluation of {LLM}s on Formal Specification Inference",
    author = "Le-Cong, Thanh  and
      Le, Bach  and
      Murray, Toby",
    editor = "Che, Wanxiang  and
      Nabende, Joyce  and
      Shutova, Ekaterina  and
      Pilehvar, Mohammad Taher",
    booktitle = "Proceedings of the 63rd Annual Meeting of the Association for Computational Linguistics (Volume 1: Long Papers)",
    month = jul,
    year = "2025",
    address = "Vienna, Austria",
    publisher = "Association for Computational Linguistics",
    url = "https://aclanthology.org/2025.acl-long.1068/",
    doi = "10.18653/v1/2025.acl-long.1068",
    pages = "21991--22014",
    ISBN = "979-8-89176-251-0",
}

@InProceedings{10.1007/978-3-031-65630-9_16,
author="Wen, Cheng
and Cao, Jialun
and Su, Jie
and Xu, Zhiwu
and Qin, Shengchao
and He, Mengda
and Li, Haokun
and Cheung, Shing-Chi
and Tian, Cong",
editor="Gurfinkel, Arie
and Ganesh, Vijay",
title="Enchanting Program Specification Synthesis by Large Language Models Using Static Analysis and Program Verification",
booktitle="Computer Aided Verification",
year="2024",
publisher="Springer Nature Switzerland",
address="Cham",
pages="302--328",
isbn="978-3-031-65630-9"
}

@misc{zheng2025livecodebenchproolympiadmedalists,
      title={LiveCodeBench Pro: How Do Olympiad Medalists Judge LLMs in Competitive Programming?}, 
      author={Zihan Zheng and Zerui Cheng and Zeyu Shen and Shang Zhou and Kaiyuan Liu and Hansen He and Dongruixuan Li and Stanley Wei and Hangyi Hao and Jianzhu Yao and Peiyao Sheng and Zixuan Wang and Wenhao Chai and Aleksandra Korolova and Peter Henderson and Sanjeev Arora and Pramod Viswanath and Jingbo Shang and Saining Xie},
      year={2025},
      eprint={2506.11928},
      archivePrefix={arXiv},
      primaryClass={cs.SE},
      url={https://arxiv.org/abs/2506.11928}, 
}

@misc{deepseekai2025deepseekv32pushingfrontieropen,
      title={DeepSeek-V3.2: Pushing the Frontier of Open Large Language Models}, 
      author={DeepSeek-AI and Aixin Liu and Aoxue Mei and Bangcai Lin and others},
      year={2025},
      eprint={2512.02556},
      archivePrefix={arXiv},
      primaryClass={cs.CL},
      url={https://arxiv.org/abs/2512.02556}, 
}

@misc{yang2025qwen3technicalreport,
      title={Qwen3 Technical Report}, 
      author={An Yang and Anfeng Li and Baosong Yang and Beichen Zhang and others},
      year={2025},
      eprint={2505.09388},
      archivePrefix={arXiv},
      primaryClass={cs.CL},
      url={https://arxiv.org/abs/2505.09388}, 
}

@misc{qwq32b,
    title = {QwQ-32B: Embracing the Power of Reinforcement Learning},
    url = {https://qwenlm.github.io/blog/qwq-32b/},
    author = {Qwen Team},
    month = {March},
    year = {2025}
}

@misc{openai2025gptoss120bgptoss20bmodel,
      title={gpt-oss-120b \& gpt-oss-20b Model Card}, 
      author={OpenAI},
      year={2025},
      eprint={2508.10925},
      archivePrefix={arXiv},
      primaryClass={cs.CL},
      url={https://arxiv.org/abs/2508.10925}, 
}

@misc{alhanahnah2025empiricalevaluationpretrainedlarge,
      title={An Empirical Evaluation of Pre-trained Large Language Models for Repairing Declarative Formal Specifications}, 
      author={Mohannad Alhanahnah and Md Rashedul Hasan and Lisong Xu and Hamid Bagheri},
      year={2025},
      eprint={2404.11050},
      archivePrefix={arXiv},
      primaryClass={cs.SE},
      url={https://arxiv.org/abs/2404.11050}, 
}

@inproceedings{yu-etal-2025-utboost,
    title = "{UTB}oost: Rigorous Evaluation of Coding Agents on {SWE}-Bench",
    author = "Yu, Boxi  and
      Zhu, Yuxuan  and
      He, Pinjia  and
      Kang, Daniel",
    editor = "Che, Wanxiang  and
      Nabende, Joyce  and
      Shutova, Ekaterina  and
      Pilehvar, Mohammad Taher",
    booktitle = "Proceedings of the 63rd Annual Meeting of the Association for Computational Linguistics (Volume 1: Long Papers)",
    month = jul,
    year = "2025",
    address = "Vienna, Austria",
    publisher = "Association for Computational Linguistics",
    url = "https://aclanthology.org/2025.acl-long.189/",
    doi = "10.18653/v1/2025.acl-long.189",
    pages = "3762--3774",
    ISBN = "979-8-89176-251-0",
}

@inproceedings{10.5555/3780338.3780368,
author = {Ahmed, Toufique and Ganhotra, Jatin and Pan, Rangeet and Shinnar, Avraham and Sinha, Saurabh and Hirzel, Martin},
title = {Otter: generating tests from issues to validate SWE patches},
year = {2026},
publisher = {JMLR.org},
booktitle = {Proceedings of the 42nd International Conference on Machine Learning},
articleno = {30},
numpages = {20},
location = {Vancouver, Canada},
series = {ICML'25}
}

@inproceedings{10.5555/3737916.3740517,
author = {M\"{u}ndler, Niels and M\"{u}ller, Mark Niklas and He, Jingxuan and Vechev, Martin},
title = {SWT-bench: testing and validating real-world bug-fixes with code agents},
year = {2024},
isbn = {9798331314385},
publisher = {Curran Associates Inc.},
address = {Red Hook, NY, USA},
booktitle = {Proceedings of the 38th International Conference on Neural Information Processing Systems},
articleno = {2601},
numpages = {31},
location = {Vancouver, BC, Canada},
series = {NIPS '24}
}

@article{10.1145/3747588,
author = {Jiang, Juyong and Wang, Fan and Shen, Jiasi and Kim, Sungju and Kim, Sunghun},
title = {A Survey on Large Language Models for Code Generation},
year = {2026},
issue_date = {February 2026},
publisher = {Association for Computing Machinery},
address = {New York, NY, USA},
volume = {35},
number = {2},
issn = {1049-331X},
url = {https://doi.org/10.1145/3747588},
doi = {10.1145/3747588},
journal = {ACM Trans. Softw. Eng. Methodol.},
month = jan,
articleno = {58},
numpages = {72}
}

@article{10.1145/3796507,
author = {Liu, Junwei and Wang, Kaixin and Chen, Yixuan and Peng, Xin and Chen, Zhenpeng and Zhang, Lingming and Lou, Yiling},
title = {Large Language Model-Based Agents for Software Engineering: A Survey},
year = {2026},
publisher = {Association for Computing Machinery},
address = {New York, NY, USA},
issn = {1049-331X},
url = {https://doi.org/10.1145/3796507},
doi = {10.1145/3796507},
note = {Just Accepted},
journal = {ACM Trans. Softw. Eng. Methodol.},
month = mar
}

@ARTICLE{161279,
  author={Meyer, B.},
  journal={Computer}, 
  title={Applying 'design by contract'}, 
  year={1992},
  volume={25},
  number={10},
  pages={40-51},
  doi={10.1109/2.161279}}

@InProceedings{10.1007/978-3-030-29026-9_22,
author="Kamburjan, Eduard",
editor="Cerrito, Serenella
and Popescu, Andrei",
title="Behavioral Program Logic",
booktitle="Automated Reasoning with Analytic Tableaux and Related Methods",
year="2019",
publisher="Springer International Publishing",
address="Cham",
pages="391--408",
isbn="978-3-030-29026-9"
}

\appendix
\clearpage

\section{Error Taxonomy and Failure Modes in Generated Specifications}
\label{sec-error-taxonomy}
To better understand failure modes of generated specifications, we conduct a fine-grained error analysis.

\subsection{\benchname{}-Func}
These errors arise from runtime failures encountered when executing the generated specifications against the provided test cases. We categorize errors into seven types in \benchname{}-Func.

\paragraph{Type Violations.}
Incorrect data type constraints.
\textit{Example:} \texttt{hours[1] must be an int, got <class 'str'>}.

\paragraph{Range Violations.}
Incorrect numeric bounds.
\textit{Example:} \texttt{Element at index 26 = 134217727 out of allowed range [0, 1e8]}.

\paragraph{Structural Violations.}
Incorrect structural constraints.
\textit{Example:} \texttt{Row 4 length must be equal to n}.

\paragraph{Relational Violations.}
Incorrect relationships among variables.
\textit{Example:} \texttt{stones must be strictly increasing (7 >= 6)}.

\paragraph{Output Mismatches.}
Incorrect return values or postconditions.
\textit{Example:} \texttt{Incorrect output: expected 6520, got 0}.

\paragraph{Interface Errors.}
Incorrect function signature.
\textit{Example:} \texttt{postconditions() takes 2 positional arguments but 4 were given}.

\paragraph{Runtime Errors.}
Syntactically invalid expressions.
\textit{Example:} \texttt{invalid syntax (line 119)}.

\begin{table*}[tb]
\centering
\begin{tabular}{lcc}
\toprule
\textbf{Error Type} & \textbf{GPT-5-mini} & \textbf{GPT-OSS-120B} \\
\midrule
Output Mismatches     & 21.5\% & 23.7\% \\
Relational Violations & 5.6\%  & 6.2\%  \\
Type Violations       & 24.9\% & 31.8\% \\
Range Violations      & 23.3\% & 22.5\% \\
Structural Violations & 8.3\%  & 10.3\% \\
Interface Errors      & 16.2\% & 4.8\%  \\
Runtime Errors        & 0.2\%  & 0.6\%  \\
\bottomrule
\end{tabular}
\caption{Distribution of error categories among failed function-level test cases.}
\label{tab-error-distribution-func}
\end{table*}

Table~\ref{tab-error-distribution-func} reports the distribution of error types among failed function-level test cases for a representative closed-source model (GPT-5-mini) and an open-source model (GPT-OSS-120B).

Both models exhibit high proportions of Type and Range Violations (48--54\% combined), indicating that many failures stem from incorrectly specified input constraints rather than flawed algorithmic reasoning. This suggests that the accurate definition of constraints remains a key challenge.

\subsection{\benchname{}-Repo}
These errors arise from runtime failures encountered when executing the generated specifications against the test cases. We categorize errors into eight types in \benchname{}-Repo.

\paragraph{Syntax Errors.}
The specification cannot be parsed due to invalid syntax.  
\textit{Example:} \texttt{SyntaxError: '(' was never closed}.

\paragraph{Dependency Resolution.}
Required imports or symbols are missing or undefined.  
\textit{Example:} \texttt{NameError: name 'ValidationError' is not defined}.

\paragraph{Signature Mismatch.}
The specification uses an incompatible function or method signature.  
\textit{Example:} \texttt{wrapped\_modify\_field\_list() got an unexpected keyword argument 'suppress\_rtype'}.

\paragraph{Execution Environment Failures.}
Errors are caused by the testing or instrumentation environment.  
\textit{Example:} \texttt{No test output found; the test suite crashed during setup}.

\paragraph{Untriggered Instrumentation.}
Instrumentation never executes during testing, indicating it was not correctly applied despite no runtime errors.

\paragraph{Over-restrictive Specifications.}
The specification is stricter than the actual valid behavior, rejecting legitimate outputs.  
\textit{Example:} \texttt{assert isinstance(r, (Quantity, np.ndarray, type(None)))}.
This assertion excludes valid return types (e.g., floats from \texttt{np.frexp}).

\paragraph{Abstraction-violating Specifications.}
The specification depends on internal implementation details instead of the externally observable behavior.
\textit{Example:} \texttt{assert not self.\_frameon or self.patch.get\_visible() == False}.
It enforces a particular internal implementation (e.g., \texttt{\_frameon}, \texttt{patch.get\_visible()}), even though alternative implementations could produce the same correct observable behavior.

\paragraph{Representation-dependent Specifications.}
The specification assumes properties that depend on a particular representation.
\textit{Example:} \texttt{assert len(X) == len(y), "X and y must have same length"}.
It assumes that \texttt{len(X)} is always defined, which fails for certain data representations (e.g., sparse matrices).

\begin{table*}[t]
\centering
\begin{tabular}{lcc}
\toprule
\textbf{Error Type} & \textbf{Claude} & \textbf{DeepSeek} \\
\midrule
Syntax Errors                          & 2.7\%  & 26.4\% \\
Dependency Resolution                  & 33.4\% & 26.4\% \\
Signature Mismatch                     & 8.7\%  & 5.9\%  \\
Execution Environment Failures         & 8.0\%  & 6.2\%  \\
Untriggered Instrumentation            & 12.4\% & 11.4\% \\
Over-restrictive Specifications        & 18.7\% & 12.7\% \\
Abstraction-violating Specifications   & 8.0\%  & 7.0\%  \\
Representation-dependent Specifications& 8.0\%  & 4.1\%  \\
\bottomrule
\end{tabular}
\caption{
Distribution of failure types on \benchname{}-Repo.
Claude refers to Claude-4.5-Sonnet and DeepSeek refers to DeepSeek-V3.2.
}
\label{tab-error-distribution-repo}
\end{table*}

Table~\ref{tab-error-distribution-repo} reports the distribution of error types for a representative closed-source model (Claude-4.5-Sonnet) and an open-source model (DeepSeek-V3.2) on \benchname{}-Repo.

These results reveal distinct failure patterns. Claude-4.5-Sonnet is primarily constrained by dependency resolution issues (33.4\%) and over-restrictive specifications (18.7\%). In contrast, DeepSeek-V3.2 exhibits a more balanced distribution, with substantial contributions from syntax errors (26.4\%) and dependency resolution failures (26.4\%). Overall, these findings suggest that Claude’s errors are largely driven by specification design and dependency handling, whereas DeepSeek’s failures are more influenced by low-level syntactic issues.

\section{\benchname{}-Func and \benchname{}-Repo Examples}
\begin{figure}[!htbp]
\centering
\includegraphics[width=\columnwidth]{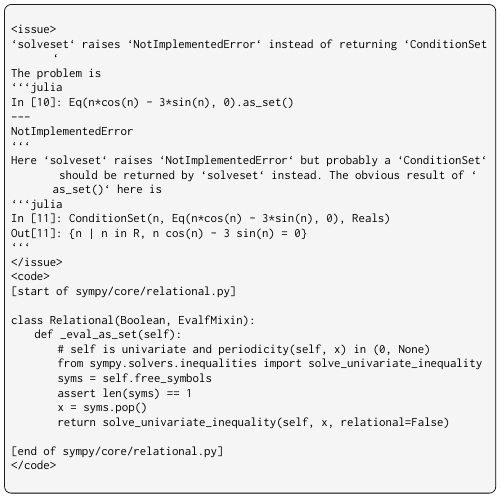}
\caption{Natural-language issue description and relevant repository code used as the input prompt for the \benchname{}-Repo example.}
\label{fig-repo-issue-prompt}
\end{figure}

Figure~\ref{fig-func-spec-example} shows an example of a natural-language problem description and its corresponding generated executable behavioral specification from \benchname{}-Func. For the repository-level example, Figure~\ref{fig-repo-issue-prompt} shows the natural-language issue description and relevant repository code used as the input prompt, and Figure~\ref{fig-repo-spec-example} shows the corresponding generated executable behavioral specification.

\begin{figure*}[tb]
    \begin{subfigure}{0.30\textwidth}
    \setlength{\FrameSep}{2pt}
    \begin{framed}
    \small
    \color{deepred}
    \texttt{Given an integer \textbf{n}, return True if n is a power of two, and False otherwise.}
    \end{framed}
    \begin{framed}
    \scriptsize
    \color{deepblue}
    \texttt{Input: n = 16\\
Output: True
}
    \end{framed}
        \caption{Problem description and example.}
        \label{fig:intent}
    \end{subfigure}
    \hspace{2mm}
    \begin{subfigure}{0.67\textwidth}
        \small
        \begin{Python}
def preconditions(n):
    assert isinstance(n, int)
    assert -2**31 <= n <= 2**31 - 1

def postconditions(n, result):
    assert isinstance(result, bool)
    expected = n > 0 and (n & (n - 1)) == 0
    assert result == expected
        \end{Python}
        \caption{Generated executable behavioral specification.}
        \label{fig:ambiguous}
    \end{subfigure}
    \caption{GPT-5 specification for \texttt{power-of-two}, a task in the revised \benchname{}-Func dataset. The code omits comments and assertion messages.}
    \label{fig-func-spec-example}
\end{figure*}

\begin{figure*}[!tbp]
\centering
\small
\begin{Python}
import pytest
from sympy.core.relational import Relational, Equality
from sympy import Symbol, cos, sin, Eq, S
from sympy.sets import ConditionSet, Set

def pytest_sessionstart(session):
    original_eval_as_set = Relational._eval_as_set

    def wrapped_eval_as_set(self):
        precondition(self)
        try:
            result = original_eval_as_set(self)
        except NotImplementedError as exc:
            postcondition(self, None, exc)
            raise
        postcondition(self, result, None)
        return result

    Relational._eval_as_set = wrapped_eval_as_set

def precondition(self):
    """Verify preconditions before calling _eval_as_set."""
    assert isinstance(self, Relational), "self must be a Relational instance"
    assert len(self.free_symbols) == 1, "Relational._eval_as_set is only defined here for univariate relations"
    return True

def postcondition(self, result, exc):
    """
    Verify behavioral contract for _eval_as_set.
    For equations like Eq(n*cos(n) - 3*sin(n), 0), _eval_as_set should not
    raise NotImplementedError. Instead, it should return a ConditionSet.
    """
    n = Symbol('n', real=True)
    difficult_eq = Eq(n*cos(n) - 3*sin(n), 0)

    is_target_case = isinstance(self, Equality) and self == difficult_eq

    if is_target_case:
        assert exc is None, (
            "_eval_as_set should return a ConditionSet for "
            "Eq(n*cos(n) - 3*sin(n), 0), not raise NotImplementedError"
        )

        assert isinstance(result, ConditionSet), (
            f"Expected ConditionSet for unsolved equality, got {type(result)}"
        )

        assert result.sym == n, "ConditionSet should bind the symbol n"
        assert result.condition == difficult_eq, (
            "ConditionSet condition should be the original equation"
        )
        assert result.base_set == S.Reals, (
            "ConditionSet base_set should be Reals for a real symbol"
        )
    else:
        if exc is None:
            assert isinstance(result, Set), (
                f"_eval_as_set should return a Set when successful, got {type(result)}"
            )

    return True
\end{Python}
\caption{Example of a generated executable behavioral specification from natural language issue description and relevant repository code, sourced from \benchname{}-Repo.}
\label{fig-repo-spec-example}
\end{figure*}

\section{Human Audit of Generated Specifications}
\label{sec-human-audit-specs}

\subsection{\benchname{}-Func Specification Audit}
\label{sec-human-audit-func}

\subsubsection{Passed Specification Audit}

We manually audited function-level specifications that satisfied the strict automated metric by accepting all valid test cases and rejecting all invalid-input and incorrect-output cases. We used a fixed random seed to sample 50 of the 1,171 GPT-5-mini specifications in the audit pool and labeled each as meaningful, weak, or strictly trivial. The results are summarized in Table~\ref{tab-gpt5mini-func-passed-spec-audit}.

\begin{table}[t]
\centering
\small
\begin{tabular}{p{0.43\linewidth}rr}
\toprule
\textbf{Human label} & \textbf{Count} & \textbf{Ratio} \\
\midrule
Meaningful / non-trivial & 50 & 100\% \\
Weak / partial but problem-related & 0 & 0\% \\
Strictly trivial / degenerate & 0 & 0\% \\
\bottomrule
\end{tabular}
\caption{Human audit of 50 randomly sampled passing GPT-5-mini function-level specifications.}
\label{tab-gpt5mini-func-passed-spec-audit}
\end{table}

We label function-level specifications using three categories: meaningful, weak / partial, and strictly trivial.  A strictly trivial function-level specification is an always-true, type-only, range-only, non-null-only, or no-op specification that does not encode the required input-output relation.  A weak / partial specification checks a property related to the problem but does not characterize the full answer.  In the sampled passed function-level specifications, we find no strictly trivial or weak specifications.  This is consistent with the \benchname{}-Func evaluation design: if a postcondition only checks type, range, non-nullness, or other shallow properties, it is likely to accept many incorrect outputs and therefore fail the completeness criterion.

\paragraph{Meaningful specification examples.}
For \texttt{shortest-bridge}, the generated specification contains both a non-trivial precondition and a semantic postcondition.  The following precondition excerpt checks that the binary grid contains exactly two four-directionally connected islands:

\begin{Python}
visited = [[False] * n for _ in range(n)]
islands = 0
for i in range(n):
    for j in range(n):
        if grid[i][j] == 1 and not visited[i][j]:
            islands += 1
            stack = [(i, j)]
            visited[i][j] = True
            while stack:
                x, y = stack.pop()
                for dx, dy in ((1, 0), (-1, 0), (0, 1), (0, -1)):
                    nx, ny = x + dx, y + dy
                    if 0 <= nx < n and 0 <= ny < n and \
                       not visited[nx][ny] and grid[nx][ny] == 1:
                        visited[nx][ny] = True
                        stack.append((nx, ny))
assert islands == 2
\end{Python}

The corresponding postcondition checks the returned number of flips by expanding outward from the first island until reaching the second island:

\begin{Python}
while q:
    x, y, d = q.popleft()
    for dx, dy in ((1, 0), (-1, 0), (0, 1), (0, -1)):
        nx, ny = x + dx, y + dy
        if 0 <= nx < n and 0 <= ny < n and not visited[nx][ny]:
            if grid[nx][ny] == 1:
                expected = d
                q.clear()
                break
            visited[nx][ny] = True
            q.append((nx, ny, d + 1))
assert output == expected
\end{Python}

This example illustrates that passed specifications need not merely recompute the whole solution in an opaque way: the precondition checks a semantic validity property of the input, while the postcondition verifies the returned distance through an independent graph-search characterization.

For \texttt{insert-greatest-common-divisors-}\allowbreak\texttt{in-linked-list}, the specification checks the structure of a returned linked list rather than only checking that a list node is returned.  It extracts the original and output values, verifies that the output length is correct, preserves all original nodes at even positions, and checks that each inserted odd-position value equals the greatest common divisor of its neighbors:

\begin{Python}
orig = _extract_values(head)
out = _extract_values(output)
expected_len = 2 * len(orig) - 1
assert len(out) == expected_len

for i in range(expected_len):
    if i % 2 == 0:
        assert out[i] == orig[i // 2]
    else:
        left = orig[i // 2]
        right = orig[i // 2 + 1]
        assert out[i] == math.gcd(left, right)
\end{Python}

Together, these examples show that passed function-level specifications often combine meaningful input constraints with output behavior checks, rather than merely asserting superficial type or range properties.

\subsubsection{Only-Correct and Only-Complete Cases}
\label{sec-func-partial-specs}

We also inspected non-passing function-level outcomes to understand why otherwise plausible specifications fail the strict metric. A correct-only specification accepts all valid inputs and correct outputs, but fails to reject at least one invalid input or incorrect output. A complete-only specification rejects all available invalid or incorrect cases, but is too restrictive for at least one valid input or correct output.

\paragraph{Only-correct specification.}
For \texttt{valid-arrangement-of-pairs}, the generated specification checks a certificate-style output for an Eulerian path.  The postcondition verifies that the proposed arrangement uses exactly the same multiset of directed pairs and that adjacent pairs are connected:

\begin{Python}
assert len(result) == len(pairs)

for i in range(1, len(result)):
    prev = result[i - 1]
    cur = result[i]
    assert prev[1] == cur[0]

from collections import Counter
in_counts = Counter(tuple(p) for p in pairs)
out_counts = Counter(tuple(p) for p in result)
assert in_counts == out_counts
\end{Python}

The specification received a correct-only label because it accepted a different valid arrangement that the oracle had labeled as a negative output.

\paragraph{A missing input constraint.}
For \emph{Minimum Swaps to Make Sequences Increasing}, GPT-5-mini's
postcondition recomputes the minimum swap count by dynamic programming
and rejects all 23 collected incorrect outputs. Its precondition checks
types, lengths, and value ranges. It accepts the pair of arrays
\texttt{[1,2,3,4,5,6]} and \texttt{[6,5,4,3,2,1]}, although the task
requires that both arrays can be made strictly increasing by swapping
elements at matching indices. Each swap preserves the sum at its index,
and all six sums are equal to 7. If both arrays were strictly increasing,
these sums would also be strictly increasing. This input therefore
violates the task guarantee, and the specification fails the precondition
completeness check.

\paragraph{Only-complete specification.}
For \texttt{maximum-binary-tree}, the generated specification is complete against the available incorrect outputs, but it rejects correct executions because of a representation-dependent output assumption.  The postcondition assumes that the output is a level-order list serialization:

\begin{Python}
preconditions(nums)
assert isinstance(output, list), \
    "output must be a list (level-order serialization)"
\end{Python}

However, the evaluation harness passes the correct output as a \texttt{TreeNode} object.  Thus the specification rejects valid behavior even though it can still reject the provided incorrect outputs.  This example shows that complete-only failures can arise from over-restrictive interface assumptions, not only from misunderstanding the algorithmic property.

\subsection{\benchname{}-Repo Specification Audit}
\label{sec-human-audit-repo}

To complement the automated execution-based metrics, we manually inspect representative repository-level specifications generated by Claude-4.5-Sonnet.  The audit focuses on whether the specification captures issue-specific behavior, whether it is degenerate or non-discriminative, and whether failures arise from semantic over-restriction or from instrumentation and dependency issues.  Correctness and completeness are computed using the same protocol as the main evaluation: correctness requires accepting the fixed-version trigger and regression tests, while completeness requires rejecting the buggy-version trigger tests.

\subsubsection{Passed Specification Audit}

We first audit only specifications that pass the automated repository-level metric.  Concretely, a specification is included in this audit only if it both accepts the fixed-version trigger and regression tests and rejects the buggy-version trigger tests, with the generated hook actually triggered during execution.  Claude-4.5-Sonnet has 101 such passed repository-level specifications.  We randomly sample 50 of them using a fixed seed and manually label whether the passed specification is meaningful, weak, or strictly trivial.  The results are summarized in Table~\ref{tab-claude-passed-spec-audit}.

\begin{table}[!htbp]
\centering
\small
\begin{tabular}{p{0.43\linewidth}rr}
\toprule
\textbf{Human label} & \textbf{Count} & \textbf{Ratio} \\
\midrule
Meaningful / non-trivial & 44 & 88\% \\
Weak / partial but issue-related & 6 & 12\% \\
Strictly trivial / degenerate & 0 & 0\% \\
\bottomrule
\end{tabular}
\caption{Human audit of 50 randomly sampled passed Claude-4.5-Sonnet specifications on \benchname{}-Repo.}
\label{tab-claude-passed-spec-audit}
\end{table}

We label a specification as \emph{strictly trivial} if it is an always-true, type-only, non-null-only, or no-op wrapper that does not encode issue-specific behavior.  We label a specification as \emph{weak / partial} if it is related to the issue but only checks a shallow or indirect property, such as the presence of an option or the absence of an exception.  The audit finds no strictly trivial passed specifications in the sample.  Most passed specifications are meaningful and issue-specific, while a smaller fraction remain partial and under-specified.

\paragraph{Meaningful specification examples.}
Although many repository-level specifications encode the discriminative behavior in postconditions, we also find passed specifications where the precondition is useful because it records the pre-state needed for a later behavioral check.  For \texttt{django\_\_django-12273}, the issue concerns resetting the primary key of a child model in multi-table inheritance.  The precondition captures the parent-link field values before \texttt{\_set\_pk\_val} runs, so that the postcondition can verify that they are reset when the primary key is set to \texttt{None}:

\begin{Python}
if value is None and hasattr(self._meta, 'parents'):
    self._pk_reset_context = {
        'has_parents': len(self._meta.parents) > 0,
        'parent_links': {
            field.attname: getattr(self, field.attname, None)
            for parent, field in self._meta.parents.items()
            if field is not None
        }
    }
\end{Python}

Similarly, \texttt{django\_\_django-11999} targets a regression where \texttt{Field.contribute\_to\_class} overwrites a user-defined \texttt{get\_FOO\_display()} method.  The precondition records whether the class already defines a custom display method before field contribution:

\begin{Python}
display_method_name = 'get_%s_display' % name
has_custom_method = (
    hasattr(cls, display_method_name) and
    display_method_name in cls.__dict__
)
field._had_custom_display_method = has_custom_method
if has_custom_method:
    field._original_display_method = getattr(cls, display_method_name)
\end{Python}

These examples show that repository-level preconditions are most informative when they snapshot mutable library or framework state before the wrapped call, rather than merely checking argument types.

For \texttt{sympy\_\_sympy-11618}, the issue concerns distance computation between points with different dimensionalities.  Claude generates a postcondition that reconstructs the expected distance by padding the shorter point with zeros before comparing against the returned value:

\begin{Python}
max_len = max(len(self_args), len(p_args))
self_padded = list(self_args) + [0] * (max_len - len(self_args))
p_padded = list(p_args) + [0] * (max_len - len(p_args))
expected_sum = sum((a - b)**2 for a, b in zip(self_padded, p_padded))
expected = sqrt(expected_sum)
assert result.equals(expected) or result == expected
\end{Python}

This specification is non-trivial because it expresses the input-output relation induced by the issue.  It passes both correctness and completeness: it accepts fixed behavior and rejects buggy behavior that computes distance without the required dimensional padding.

Another meaningful passed example is \texttt{astropy\_\_astropy-14309}, where the issue concerns FITS identification.  The generated specification checks not only that \texttt{is\_fits} returns a boolean, but also that FITS file extensions and FITS object inputs are accepted while non-FITS paths are rejected:

\begin{Python}
assert isinstance(result, bool)

if filepath is not None and fileobj is None and len(args) == 0:
    if not filepath.lower().endswith(
        (".fits", ".fits.gz", ".fit", ".fit.gz", ".fts", ".fts.gz")
    ):
        assert result is False

if len(args) > 0 and isinstance(args[0], (HDUList, TableHDU, BinTableHDU, GroupsHDU)):
    assert result is True

if filepath is not None and filepath.lower().endswith(
    (".fits", ".fits.gz", ".fit", ".fit.gz", ".fts", ".fts.gz")
):
    assert result is True
\end{Python}

This directly encodes the issue-level input-output behavior rather than only checking executability.

\paragraph{Weak but passed specification.}
The audit also identifies passed specifications that are issue-related but incomplete as behavioral descriptions.  For example, \texttt{django\_\_django-12143} targets a bug where admin formset prefixes containing regex metacharacters must be escaped before extracting edited object primary keys.  The generated specification checks that matching POST keys are included in the result, but its false-positive check is vacuous:

\begin{Python}
for value in result:
    found = False
    for key, post_value in request.POST.items():
        if expected_pattern.match(key) and post_value == value:
            found = True
            break
    assert found or True, \
        f"Result value {value} should correspond to a POST key matching pattern"
\end{Python}

This specification is not strictly trivial: it constructs an escaped-prefix pattern and checks that true matches are included.  However, it is weak because the \texttt{assert found or True} branch cannot fail, so it does not fully rule out extra primary keys that do not correspond to escaped-prefix matches.  We therefore label it as weak / partial rather than meaningful.

\subsubsection{Partial and Failed Execution Cases}

We separately inspect three non-passing outcomes because they diagnose different ways in which repository-level specifications can fail.  A correct-only specification accepts fixed behavior but does not reject buggy behavior.  A complete-only specification rejects buggy behavior but is too strict for valid fixed behavior.  A failed hook or injection case cannot reliably execute the intended specification in the repository test harness.

\paragraph{Weak non-discriminative specification.}
For \texttt{psf\_\_requests-1724}, the issue is about converting request methods to native strings before constructing a request.  The generated postcondition largely checks only that execution reaches the end of \texttt{Session.request}; on Python 3, the branch containing the intended Unicode-specific check is inactive:

\begin{Python}
def precondition(method):
    assert isinstance(method, str), "method must be str in Python 3"
    return True

def postcondition(original_method, session_obj):
    if sys.version_info[0] == 2:
        pass
    return True
\end{Python}

This example clarifies how we identify non-discriminative specifications.  The spec is executable and accepts the fixed repository, but it does not encode a property that distinguishes the fixed behavior from the buggy behavior under the evaluated runtime.  Accordingly, it is counted as correct-only.

\paragraph{Over-restrictive specification.}
For \texttt{astropy\_\_astropy-13977}, the issue concerns returning \texttt{NotImplemented} for unsupported \texttt{Quantity.\_\_array\_ufunc\_\_} interactions.  Claude captures part of the intended behavior but introduces an overly narrow type constraint for tuple-valued results:

\begin{Python}
if isinstance(result, (tuple, list)):
    for r in result:
        assert isinstance(r, (Quantity, np.ndarray, type(None))), \
            "Tuple results must contain Quantity, ndarray, or None"
    return True
\end{Python}

The fixed repository can validly return scalar values inside tuple outputs, such as \texttt{0.5} from \texttt{np.frexp}.  The generated assertion therefore rejects valid fixed behavior.  This illustrates a complete-only failure mode: the model writes a plausible local invariant that is stricter than the library's actual public behavior.

\paragraph{Failed hook / target-resolution case.}
For \texttt{sympy\_\_sympy-16597}, the issue concerns assumption propagation: a symbol declared with \texttt{even=True} or \texttt{odd=True} should also be inferred to be finite.  The generated specification attempts to wrap the assumption-fact generator and then check the intended behavior:

\begin{Python}
from sympy.assumptions import ask as ask_module
original_get_known_facts = ask_module.get_known_facts

def wrapped_get_known_facts():
    result = original_get_known_facts()
    postcondition_known_facts(result)
    return result

m = Symbol('m', even=True)
assert m.is_finite == True
\end{Python}

However, the hook fails before the behavioral checks can run.  In this environment, \texttt{from sympy.assumptions import ask as ask\_module} binds \texttt{ask\_module} to the \texttt{ask} function rather than to the \texttt{sympy.assumptions.ask} module.  Consequently, \texttt{pytest\_sessionstart} raises \texttt{AttributeError: 'function' object has no attribute 'get\_known\_facts'}.  This case is not a trivial semantic specification; rather, it shows that repository-level specifications can fail when the generated hook resolves or wraps the wrong project object.

\section{Repository-level Specification Injection}
\label{sec-repo-spec-injection}

Most repositories in SWE-bench Verified use \texttt{pytest} for testing. During execution, \texttt{pytest} automatically loads \texttt{conftest.py} files from test directories and their parent directories. If a \texttt{conftest.py} defines the \texttt{pytest\_sessionstart} hook, it is executed once at the beginning of the test session, before any test cases run.

To enable automated validation of generated specifications, the model produces a \texttt{conftest.py} file containing (i) the specification functions (\texttt{preconditions} and \texttt{postconditions}) and (ii) a \texttt{pytest\_sessionstart} function that installs a hook on the target function. The hook wraps the original function so that \texttt{preconditions} execute before the function call and \texttt{postconditions} execute after it returns, effectively replacing the original function at runtime.

During testing, the generated \texttt{conftest.py} is placed at the repository root (or appended to an existing one). Because \texttt{pytest} automatically invokes \texttt{pytest\_sessionstart} at startup, the specifications are activated during test execution.

\paragraph{Django.}
Django does not use \texttt{pytest}; instead, it relies on a custom test runner implemented in \texttt{tests/runtests.py}. Before executing tests, this script calls \texttt{setup\_run\_tests} for initialization. To integrate our mechanism, we remove the \texttt{import pytest} statement from the generated \texttt{conftest.py} and append its contents to the \texttt{setup\_run\_tests} function, ensuring that the same hooking mechanism is applied before tests start.

\paragraph{SymPy.}
SymPy originally runs tests using \texttt{bin/test}. To support our injection mechanism, we switch the test command to \texttt{pytest}, since SymPy’s test cases are compatible with \texttt{pytest}. This enables the same function-wrapping strategy for executing \texttt{preconditions} and \texttt{postconditions}. We also update the test-output parsing logic to match the \texttt{pytest} format so that results are recorded correctly.

\section{Quality Assurance}
\label{sec-quality-assurance}

To ensure the reliability of \benchname{}-Func, we incorporate multiple quality assurance measures throughout the benchmark construction process. 

First, to further improve evaluation reliability, we substantially expand the number of test cases per problem. The revised benchmark has 215.5 checks per problem on average, counting precondition and postcondition checks separately. This expanded test suite provides stronger evaluation signals compared with prior benchmarks, which typically include far fewer test cases per problem.

We use the official online judge to label generated inputs and retain inputs with deterministic execution outcomes. We deduplicate inputs that differ only in formatting. Output comparisons use the judge's reference, so valid alternatives can receive negative labels. The revised Func split excludes the 147 candidate multi-answer problems identified by the description-based screen.

We also conduct quantitative coverage analysis to assess the diversity and effectiveness of the generated tests. Among the 2{,}346 retained tasks with coverage reports, the reference solutions have mean statement coverage of 96.4\% and mean branch coverage of 93.8\%. The retained \texttt{reverse-bits} task has no available coverage report.

For evaluating postcondition completeness, incorrect output cases are constructed using candidate solutions generated by LLMs with diverse capability levels. The resulting incorrect test cases reflect realistic error distributions, providing meaningful signals.

For \benchname{}-Repo, we augment the original test suites of SWE-bench Verified with additional test cases from UTBoost. These tests address known coverage gaps, ensuring each instance provides sufficiently comprehensive and rigorous evaluation.

\section{Fraction-Based Evaluation Metrics}
\label{sec-fraction-metric}
\definecolor{scbf}{RGB}{235,242,255}
\definecolor{scbr}{RGB}{255,239,230}
\definecolor{groupPink}{RGB}{255,230,230}
\definecolor{groupGreen}{RGB}{220,255,220}

\begin{table*}[tb]
\centering
\small
\begin{tabular}{l c c c c c c c}

\toprule
\multirow{2}{*}{\textbf{Model}} & \multirow{2}{*}{\textbf{Reasoning}}
& \multicolumn{3}{>{\columncolor{scbf}}c}{\textbf{\benchname{}-Func}}
& \multicolumn{3}{>{\columncolor{scbr}}c}{\textbf{\benchname{}-Repo}} \\
\cmidrule(lr){3-5} \cmidrule(lr){6-8}
 &
 & \cellcolor{scbf}\textbf{Correct} 
 & \cellcolor{scbf}\textbf{Complete} 
 & \cellcolor{scbf}\textbf{Pass} 
 & \cellcolor{scbr}\textbf{Correct} 
 & \cellcolor{scbr}\textbf{Complete} 
 & \cellcolor{scbr}\textbf{Pass} \\
\midrule

\rowcolor{groupGreen}
\multicolumn{8}{c}{\textit{Open-Weight Models}} \\
\addlinespace[2pt]

\multirow{2}{*}{\textbf{Qwen3-0.6B}} & \ding{55} & 13.4 & 55.7 & 7.0 & 0.0 & 0.0 & 0.0 \\
          & \ding{51} & 7.5 & 63.3 & 4.0 & 0.0 & 0.0 & 0.0 \\
\addlinespace[2pt]

\multirow{2}{*}{\textbf{Qwen3-1.7B}} & \ding{55} & 13.5 & 78.1 & 10.6 & 0.4 & 0.2 & 0.3 \\
           & \ding{51} & 20.2 & 75.6 & 16.8 & 0.0 & 0.0 & 0.0 \\
\addlinespace[2pt]

\multirow{2}{*}{\textbf{Qwen3-4B}} & \ding{55} & 57.2 & 74.2 & 45.5 & 0.7 & 1.2 & 0.9 \\
         & \ding{51} & 84.3 & 86.4 & 78.5 & 1.6 & 1.2 & 1.4 \\
\addlinespace[2pt]

\multirow{2}{*}{\textbf{Qwen3-8B}} & \ding{55} & 62.4 & 75.4 & 51.1 & 7.8 & 10.0 & 8.8 \\
         & \ding{51} & 83.7 & 83.4 & 74.0 & 6.6 & 7.3 & 7.0 \\
\addlinespace[2pt]

\multirow{2}{*}{\textbf{Qwen3-14B}} & \ding{55} & 77.3 & 78.5 & 66.1 & 16.5 & 17.6 & 17.0 \\
          & \ding{51} & 89.6 & 90.2 & \underline{84.5} & 19.8 & 17.9 & 18.8 \\
\addlinespace[2pt]

\multirow{2}{*}{\textbf{Qwen3-32B}} & \ding{55} & 81.2 & 79.7 & 70.5 & 9.0 & 8.5 & 8.7 \\
          & \ding{51} & 89.6 & 88.5 & 82.6 & 21.5 & 15.8 & 18.2 \\
\addlinespace[2pt]

\textbf{QwQ-32B} & \ding{51} & \textbf{90.7} & 87.7 & 83.1 & 25.3 & 25.9 & 25.6 \\
\textbf{GPT-OSS-20B} & \ding{51} & 82.7 & \underline{94.8} & 81.2 & \underline{36.6} & \underline{32.6} & 34.5 \\
\textbf{GPT-OSS-120B} & \ding{51} & \underline{90.4} & \textbf{95.4} & \textbf{88.5} & 34.8 & \textbf{37.0} & \textbf{35.9} \\
\textbf{DeepSeek-V3.2} & \ding{55} & 86.1 & 82.7 & 76.4 & \textbf{41.6} & 30.8 & \underline{35.4} \\
\midrule

\rowcolor{groupPink}
\multicolumn{8}{c}{\textit{Proprietary Models}} \\
\addlinespace[2pt]

\textbf{Gemini-2.5-Flash} & \ding{51} & 92.4 & 90.8 & 87.5 & 26.3 & 18.1 & 21.5 \\
\textbf{Gemini-2.5-Pro} & \ding{51} & \textbf{94.9} & \underline{95.4} & \textbf{92.6} & \underline{58.1} & 53.9 & \underline{55.9} \\
\textbf{GPT-5-mini} & \ding{51} & 88.4 & \textbf{96.5} & 86.8 & 45.0 & 46.4 & 45.7 \\
\textbf{GPT-5} & \ding{51} & 88.5 & \underline{95.4} & 87.5 & 52.0 & \textbf{59.4} & 55.5 \\
\textbf{Claude-4.5-Sonnet} & \ding{51} & \underline{94.8} & 92.8 & \underline{91.3} & \textbf{67.9} & \underline{57.9} & \textbf{62.5} \\

\bottomrule
\end{tabular}
\caption{Fraction-based results on the 2{,}347 Func tasks and 500 Repo issues (percentages).}
\label{tab-speccodebench-fraction-results}
\end{table*}

In strict evaluation, a specification satisfies correctness if it accepts all collected valid tests and completeness if it rejects all collected invalid tests. Strict pass requires satisfying both criteria.

To provide a more fine-grained characterization of model behavior, we additionally report fraction-based metrics that measure the extent to which a specification satisfies correctness and completeness over the test suite.

Specifically, we define:

\begin{itemize}
    \item \textbf{Correctness}: the fraction of correct test cases that are accepted by the generated specification.

    \item \textbf{Completeness}: the fraction of incorrect test cases that are rejected by the generated specification.

    \item \textbf{Pass Rate}: the harmonic mean of Correctness and Completeness, defined as
    \begin{equation}
        \text{Pass} =
        2 \cdot
        \frac{\text{Correctness} \cdot \text{Completeness}}
        {\text{Correctness} + \text{Completeness}}
    \end{equation}
\end{itemize}

On Func, we compute these three metrics per task and average each metric across the 2{,}347 retained tasks. On Repo, we average correctness and completeness across the 500 issues before taking their harmonic mean. Table~\ref{tab-speccodebench-fraction-results} reports the detailed results.
Compared to the strict metrics in the main text, these fraction-based measures capture partial success and offer higher resolution, which is particularly informative on challenging settings such as \benchname{}-Repo, where many model outputs fail the strict criteria but still differ meaningfully in how many tests they satisfy. Importantly, although these graded metrics provide finer granularity, they typically preserve the same overall ranking trends and scaling behaviors observed under strict evaluation, further supporting the robustness of our main conclusions.

\section{Multi-Answer Task Exclusion}
\label{sec-multianswer-sensitivity}
\begin{table*}[tb]
\centering
\small
\setlength{\tabcolsep}{4pt}
\begin{tabular}{lrrrr}
\toprule
\multirow{2}{*}{\textbf{Model}} & \multicolumn{2}{c}{\textbf{Strict pass}} & \multicolumn{2}{c}{\textbf{Fractional pass}} \\
\cmidrule(lr){2-3}\cmidrule(lr){4-5}
 & Original & Revised & Original & Revised \\
\midrule
Qwen3-32B & 17.6 & 18.0 & 69.9 & 70.5 \\
GPT-OSS-120B & 42.5 & 44.0 & 87.9 & 88.5 \\
Gemini-2.5-Pro & 46.2 & 47.8 & 91.9 & 92.6 \\
GPT-5-mini & 47.0 & 48.6 & 86.3 & 86.8 \\
Claude-4.5-Sonnet & 40.5 & 41.8 & 90.7 & 91.3 \\
\bottomrule
\end{tabular}
\caption{Func specification pass rates (percent) on the original 2{,}494 tasks and the revised 2{,}347 tasks.}
\label{tab-multianswer-sensitivity}
\end{table*}

The description-based screen flagged 147 of the original 2{,}494 tasks (5.9\%). Of these, 103 had mismatch-based incorrect-output checks. Together the flagged tasks accounted for 11{,}329 of the original 169{,}176 incorrect-output checks (6.7\%). We excluded these tasks in full for the revised Func evaluation. Across the 21 model configurations, the largest absolute changes are 1.70 percentage points in strict pass and 0.93 points in fractional pass. Table~\ref{tab-multianswer-sensitivity} shows selected model results.

\section{Confidence Intervals and Repeated Runs}
\label{sec-bootstrap-ci}
\begin{table*}[tb]
\centering
\small
\setlength{\tabcolsep}{4pt}
\begin{tabular}{lrrrr}
\toprule
\multirow{2}{*}{\textbf{Model}} & \multicolumn{2}{c}{\textbf{Func (2{,}347 tasks)}} & \multicolumn{2}{c}{\textbf{Repo (500 issues)}} \\
\cmidrule(lr){2-3}\cmidrule(lr){4-5}
 & Strict pass & Fractional pass & Strict pass & Fractional pass \\
\midrule
Qwen3-0.6B & 0.2 [0.0, 0.4] & 7.0 [6.3, 7.7] & 0.0 [0.0, 0.0] & 0.0 [0.0, 0.0] \\
Qwen3-0.6B (thinking) & 0.3 [0.1, 0.5] & 4.0 [3.5, 4.6] & 0.0 [0.0, 0.0] & 0.0 [0.0, 0.0] \\
Qwen3-1.7B & 0.8 [0.5, 1.2] & 10.6 [9.6, 11.6] & 0.0 [0.0, 0.0] & 0.3 [0.0, 0.7] \\
Qwen3-1.7B (thinking) & 1.3 [0.9, 1.7] & 16.8 [15.5, 18.1] & 0.0 [0.0, 0.0] & 0.0 [0.0, 0.0] \\
Qwen3-4B & 5.7 [4.8, 6.6] & 45.5 [43.9, 47.0] & 0.0 [0.0, 0.0] & 0.9 [0.2, 1.6] \\
Qwen3-4B (thinking) & 16.0 [14.5, 17.5] & 78.5 [77.2, 79.8] & 0.2 [0.0, 0.6] & 1.4 [0.5, 2.3] \\
Qwen3-8B & 6.0 [5.0, 6.9] & 51.1 [49.5, 52.7] & 0.8 [0.2, 1.6] & 8.8 [6.7, 11.1] \\
Qwen3-8B (thinking) & 17.3 [15.9, 18.9] & 74.0 [72.5, 75.3] & 0.0 [0.0, 0.0] & 7.0 [5.1, 9.0] \\
Qwen3-14B & 13.9 [12.5, 15.3] & 66.1 [64.6, 67.5] & 1.2 [0.4, 2.2] & 17.0 [14.2, 19.9] \\
Qwen3-14B (thinking) & 27.3 [25.5, 29.1] & 84.5 [83.3, 85.7] & 1.2 [0.4, 2.2] & 18.8 [15.9, 21.6] \\
Qwen3-32B & 18.0 [16.5, 19.6] & 70.5 [69.1, 71.8] & 1.2 [0.4, 2.2] & 8.7 [6.6, 10.9] \\
Qwen3-32B (thinking) & 26.6 [24.8, 28.4] & 82.6 [81.3, 83.8] & 2.0 [0.8, 3.4] & 18.2 [15.2, 21.2] \\
QwQ-32B & 29.7 [27.8, 31.5] & 83.1 [81.9, 84.3] & 3.6 [2.0, 5.4] & 25.6 [22.5, 28.7] \\
GPT-OSS-20B & 39.0 [37.1, 41.0] & 81.2 [79.7, 82.5] & 3.4 [1.8, 5.0] & 34.5 [30.9, 37.9] \\
GPT-OSS-120B & 44.0 [42.0, 46.0] & 88.5 [87.4, 89.6] & 3.2 [1.8, 4.8] & 35.9 [32.3, 39.4] \\
DeepSeek-V3.2 & 24.9 [23.2, 26.7] & 76.4 [75.1, 77.7] & 6.8 [4.8, 9.0] & 35.4 [31.7, 38.9] \\
Gemini-2.5-Flash & 37.8 [35.8, 39.8] & 87.5 [86.4, 88.5] & 2.8 [1.4, 4.4] & 21.4 [18.4, 24.6] \\
Gemini-2.5-Pro & 47.8 [45.8, 49.8] & 92.6 [91.8, 93.5] & 18.2 [14.8, 21.6] & 55.9 [52.4, 59.5] \\
GPT-5-mini & 48.6 [46.6, 50.7] & 86.8 [85.6, 88.1] & 9.6 [7.2, 12.2] & 45.7 [42.3, 49.2] \\
GPT-5 & 41.6 [39.6, 43.5] & 87.5 [86.4, 88.5] & 8.6 [6.2, 11.2] & 55.5 [52.0, 58.7] \\
Claude-4.5-Sonnet & 41.8 [39.9, 43.8] & 91.3 [90.4, 92.1] & 20.2 [16.8, 23.8] & 62.5 [59.3, 65.8] \\
\bottomrule
\end{tabular}
\caption{Specification pass rates with 95\% task-bootstrap confidence intervals (percent).}
\label{tab-bootstrap-ci}
\end{table*}

We estimate task-level percentile 95\% confidence intervals using 10{,}000 bootstrap samples with replacement and seed 20260710. Each sample has the size of its benchmark split (2{,}347 Func tasks or 500 Repo issues). For Repo fractional pass, we take the harmonic mean of the resampled average correctness and completeness, using the same resampled issue indices for both. The intervals quantify finite-task sampling uncertainty conditional on the stored generations. The largest half-widths for strict and fractional pass are 2.05 and 1.59 percentage points on Func and 3.50 and 3.59 on Repo, respectively. Table~\ref{tab-bootstrap-ci} reports the pass intervals for all model configurations.

\paragraph{Repeated generation.}
To measure run-to-run variation, we sample five independent specifications per task from Qwen3-32B (non-thinking) and GPT-5-mini. Each run uses the same evaluation set of 2{,}347 retained Func tasks or 500 Repo issues. Table~\ref{tab-repeated-runs} reports the minimum and maximum strict pass rates across the five runs for each model and split.

\begin{table}[t]
\centering
\small
\begin{tabular}{lcc}
\toprule
\textbf{Model} & \textbf{Func} & \textbf{Repo} \\
\midrule
Qwen3-32B & 17.0--17.9 & 0.4--1.6 \\
GPT-5-mini & 50.4--51.7 & 10.6--15.8 \\
\bottomrule
\end{tabular}
\caption{Strict pass ranges (\%) over five generation runs per model and split.}
\label{tab-repeated-runs}
\end{table}

\section{Syntactic Injection Feasibility in Repo-Level Specification Generation}
\begin{table}[tb]
\centering
\scriptsize
\setlength{\tabcolsep}{3pt}
\resizebox{\columnwidth}{!}{%
\begin{tabular}{lcccc}
\toprule
\textbf{Model} & \textbf{Reasoning} & \textbf{Executable} & \textbf{Pass} \\
\midrule
\multirow{2}{*}{Qwen3-0.6B}  & \ding{55} & 0.0  & 0.0 \\
                            & \ding{51} & 0.0  & 0.0 \\
\multirow{2}{*}{Qwen3-1.7B}  & \ding{55} & 0.4  & 0.0 \\
                            & \ding{51} & 0.0  & 0.0 \\
\multirow{2}{*}{Qwen3-4B}    & \ding{55} & 1.0  & 0.0 \\
                            & \ding{51} & 2.8  & 0.2 \\
\multirow{2}{*}{Qwen3-8B}    & \ding{55} & 12.2 & 0.8 \\
                            & \ding{51} & 9.8  & 0.0 \\
\multirow{2}{*}{Qwen3-14B}   & \ding{55} & 28.6 & 1.2 \\
                            & \ding{51} & 30.6 & 1.2 \\
\multirow{2}{*}{Qwen3-32B}   & \ding{55} & 17.4 & 1.2 \\
                            & \ding{51} & 25.8 & 2.0 \\
QwQ-32B        & \ding{51} & 29.8 & 3.6 \\
GPT-OSS-20B    & \ding{51} & 48.2 & 3.4 \\
GPT-OSS-120B            & \ding{51} & 49.4 & 3.2 \\
DeepSeek-V3.2           & \ding{55} & 52.0 & 6.8 \\
Gemini-2.5-Flash        & \ding{51} & 43.8 & 2.8 \\
Gemini-2.5-Pro          & \ding{51} & 71.2 & 18.2 \\
GPT-5-Mini              & \ding{51} & 73.6 & 9.6 \\
GPT-5                   & \ding{51} & 77.6 & 8.6 \\
Claude-4.5-Sonnet       & \ding{51} & 82.0 & 20.2 \\
\bottomrule
\end{tabular}
}
\caption{Full experimental results showing the relationship between syntactic injection feasibility (Executable) and pass rate (Pass) on \benchname{}-Repo (values in \%).}
\label{tab-wrapper-vs-pass-full}
\end{table}

To understand the low pass rates on \benchname{}-Repo, we analyze whether failures are partly driven by the difficulty of generating specifications that are syntactically feasible to inject into the repository test harness.  Here, \emph{Executable} measures whether the generated wrapper can be installed and invoked by the harness, independent of whether the injected preconditions and postconditions are semantically correct.
Table~\ref{tab-wrapper-vs-pass-full} shows that syntactic injection feasibility is broadly associated with pass rate. For weaker models, only a small fraction of generated specifications can be injected and invoked (e.g., 9.8\% for Qwen3-8B-Thinking), resulting in near-zero pass rates. This suggests that injection feasibility is a major bottleneck for weaker models in the repository-level setting.
For stronger models such as GPT-5 and Claude-4.5-Sonnet, most generated specifications are syntactically feasible to inject (77.6\% and 82.0\%, respectively), yet overall pass rates remain limited. This suggests that, beyond injection feasibility, the remaining challenge lies in correctly understanding repository-level behavior and expressing it through accurate executable constraints.

\section{Scaffold-Given Ablation}
\label{sec-repo-scaffold-ablation}

We extracted scaffolds from existing \benchname{}-Repo generations that passed both correctness and completeness, yielding candidates for 200 issues. We replaced the precondition and postcondition bodies with placeholders and retained the imports, hook installation, wrappers, and function signatures. We excluded two issues whose scaffolds retained additional helper definitions. For each of the remaining 198 issues, we provided the scrubbed scaffold and asked the model to fill in the condition bodies. Table~\ref{tab-repo-scaffold-ablation} compares this setting with each model's original end-to-end result on the same issues.

Because selection required a passing generation, the comparison estimates the benefit of supplying a scaffold on issues for which one could be extracted. All four models improve on this subset.

\section{Specification vs. Solution Generation Pass Rates}
\label{sec-spec-vs-solution-table}
\begin{table*}[tb]
\centering
\small
\setlength{\tabcolsep}{4pt}
\begin{tabular}{lrrrrrrrr}
\toprule
\multirow{2}{*}{\textbf{Model}} & \multicolumn{2}{c}{\textbf{Func strict}} & \multicolumn{2}{c}{\textbf{Func fractional}} & \multicolumn{2}{c}{\textbf{Repo strict}} & \multicolumn{2}{c}{\textbf{Repo fractional}} \\
\cmidrule(lr){2-3}\cmidrule(lr){4-5}\cmidrule(lr){6-7}\cmidrule(lr){8-9}
 & Spec & Sol & Spec & Sol & Spec & Sol & Spec & Sol \\
\midrule
Gemini-2.5-Pro & -- & -- & -- & -- & 18.2 & 21.8 & 55.9 & 31.1 \\
GPT-5 & 41.6 & 88.2 & 87.5 & 93.8 & 8.6 & 19.2 & 55.5 & 38.6 \\
Claude-4.5-Sonnet & 41.8 & 91.3 & 91.3 & 95.4 & 20.2 & 45.8 & 62.5 & 75.5 \\
Qwen3-32B & 18.0 & 72.8 & 70.5 & 83.8 & 1.2 & 4.4 & 8.7 & 8.3 \\
Qwen3-14B & 13.9 & 67.4 & 66.1 & 79.7 & 1.2 & 7.4 & 17.0 & 28.6 \\
Qwen3-8B & 6.0 & 60.6 & 51.1 & 74.7 & 0.8 & 4.8 & 8.8 & 20.4 \\
\bottomrule
\end{tabular}
\caption{Specification (Spec) and solution (Sol) pass rates in percent on 2{,}347 Func tasks and 500 Repo issues. Repo solution rates come from our patch-generation runs.}
\label{tab-spec-vs-solution}
\end{table*}

Table~\ref{tab-spec-vs-solution} reports strict and fractional pass rates for five paired models on the same 2{,}347 \benchname{}-Func tasks. Strict Spec pass requires accepting all collected valid behaviors and rejecting all collected invalid behaviors. Strict Sol pass requires passing all stored valid solution tests. Fractional Spec pass is the task-level harmonic mean defined in Appendix~\ref{sec-fraction-metric}, and fractional Sol pass is the share of stored valid solution tests passed per task. For Qwen3-8B and Qwen3-32B, we average the ten stored solution samples within each task to estimate pass@1 and its fractional counterpart. Each other model contributes one stored solution sample per task. We then average across tasks.

The Func comparison pairs runs by task ID and uses each run's stored tests. The older Qwen3-8B and Qwen3-32B solution logs have different recorded valid-input sets from the revised dataset on 57 and 7 retained tasks, respectively.

On the 500 \benchname{}-Repo issues, models generate unified diffs without tool interaction. We score one patch per issue with the SWE-bench harness. Qwen3 comparisons pair non-thinking specification runs with default solution settings.

Strict Sol pass is the SWE-bench resolved rate, which requires the patch to apply and all \texttt{FAIL\_TO\_PASS} and \texttt{PASS\_TO\_PASS} tests to pass. Fractional Sol pass averages the share of these tests passed per issue over all 500 issues. Missing patches and application or execution failures score zero on both metrics. Repo fractional Spec pass uses the harmonic mean of aggregate correctness and completeness (Appendix~\ref{sec-fraction-metric}).

\paragraph{Function-level qualitative examples.}
We examine GPT-5 cases on \benchname{}-Func where the generated solution passes all validation tests and the specification fails by rejecting a correct output or accepting an incorrect output.

For the IP-defanging task, GPT-5 generates a solution that passes all validation tests:
\begin{Python}
# Problem: defanging-an-ip-address
class Solution:
    def defangIPaddr(self, address):
        return address.replace('.', '[.]')
\end{Python}
However, the generated postcondition rejects the correct output because it incorrectly asserts that a defanged address should contain no period character at all:
\begin{Python}
def postconditions(address, result):
    expected = address.replace('.', '[.]')
    assert isinstance(result, str)
    assert result == expected
    assert '.' not in result
\end{Python}
For example, for the input \texttt{"203.0.113.0"}, the oracle answer is \texttt{"203[.]0[.]113[.]0"}. Although this output exactly matches the required defanged form, the specification rejects it because the token \texttt{[.]} necessarily contains a period. This is a correctness failure even though the corresponding solution is correct.

For the cell-reachability task, GPT-5 generates a solution that passes all validation tests. The solution computes the Chebyshev distance between the start and target cells and handles the special case where the start and target are identical:
\begin{Python}
# Problem: determine-if-a-cell-is-reachable-at-a-given-time
class Solution:
    def isReachableAtTime(self, sx, sy, fx, fy, t):
        dist = max(abs(sx - fx), abs(sy - fy))
        if dist == 0:
            return t != 1
        return dist <= t
\end{Python}
However, the generated postcondition omits this exception and checks only whether \texttt{t >= min\_steps}:
\begin{Python}
def postconditions(sx, sy, fx, fy, t, output):
    assert isinstance(output, bool)
    min_steps = max(abs(fx - sx), abs(fy - sy))
    expected = (t >= min_steps)
    assert output == expected
\end{Python}
As a result, for the test input \texttt{(1, 2, 1, 2, 1)}, the oracle answer is \texttt{False}, but the specification accepts the incorrect output \texttt{True}. This specification is therefore incomplete even though the corresponding solution is correct.

For the infinite-array minimum-subarray task, GPT-5 again generates a solution that passes all validation tests. The following excerpt shows that the solution decomposes the target into full repetitions of \texttt{nums} and a remaining subarray target:
\begin{Python}
# Problem: minimum-size-subarray-in-infinite-array
class Solution:
    def minSizeSubarray(self, nums, target):
        total = sum(nums)
        n = len(nums)
        k = target // total
        rem = target % total
        if rem == 0:
            return k * n
        # search for the shortest subarray summing to rem in nums + nums
        return k * n + min_len if min_len != float("inf") else -1
\end{Python}
By contrast, the generated postcondition searches only inside \texttt{nums * 2}:
\begin{Python}
def postconditions(nums, target, output):
    N = len(nums)
    arr = nums * 2
    found = False
    for start in range(2 * N):
        s = 0
        for end in range(start, 2 * N):
            s += arr[end]
            if s == target:
                found = True
                break
    if output == -1:
        assert not found
\end{Python}
This finite-window check misses answers requiring many complete repetitions of the array. For example, with \texttt{nums = [1, 2, 3]} and \texttt{target = 30}, the oracle answer is \texttt{15}, but the specification accepts the incorrect output \texttt{-1} because no such subarray appears in the two-copy window. This case shows that a model can implement the algorithmic solution correctly while still failing to express the full behavioral specification.

\paragraph{Repository-level qualitative examples.}
On \benchname{}-Repo, the Qwen3-32B patches below resolve their issues in SWE-bench evaluation. The corresponding specifications from the Qwen3-32B thinking runs fail the strict specification evaluation.

For \texttt{django\_\_django-14089}, the issue asks for \texttt{OrderedSet} to support Python's \texttt{reversed()} protocol.  The generated solution is correct and passes the SWE-bench patch evaluation by adding \texttt{\_\_reversed\_\_}:
\begin{Python}
class OrderedSet:
    def __reversed__(self):
        return reversed(list(self.dict))
\end{Python}
However, the generated specification treats the buggy behavior as a precondition and then patches the class inside the specification itself:
\begin{Python}
def precondition():
    s = OrderedSet([1, 2, 3])
    try:
        reversed(s)
    except TypeError:
        pass
    else:
        assert False, "Expected TypeError before applying the patch"

def pytest_sessionstart(session):
    precondition()
    def __reversed__(self):
        return reversed(self.dict)
    OrderedSet.__reversed__ = __reversed__
    postcondition()
\end{Python}
This is not a behavioral specification of the fixed repository.  It asserts that \texttt{reversed()} should fail before the patch and mutates the repository at test time, rather than checking that the submitted implementation already provides the required behavior.  Consequently, the patch is marked as resolved, but the specification is rejected by the repository-level specification evaluation.

For \texttt{scikit-learn\_\_scikit-learn-11578}, the issue requires \texttt{\_log\_reg\_scoring\_path()} to pass the requested \texttt{multi\_class} mode into the internal \texttt{LogisticRegression} object.  The generated solution makes exactly this change:
\begin{Python}
log_reg = LogisticRegression(
    fit_intercept=fit_intercept,
    multi_class=multi_class,
)
\end{Python}
Nevertheless, the generated specification contains an ill-typed assertion:
\begin{Python}
def precondition(actual_multi, expected_multi):
    assert 'multi_class' in actual_multi, \
        "multi_class missing in LogisticRegression parameters"
    assert actual_multi == expected_multi
\end{Python}
During evaluation, \texttt{actual\_multi} is the string value \texttt{"multinomial"}, not a dictionary of constructor arguments.  Thus the specification rejects the correct execution by checking whether the substring \texttt{"multi\_class"} occurs inside \texttt{"multinomial"}.  This example again shows that a model can identify and implement the correct patch while failing to express the corresponding executable specification.

\section{Specification-Guided Re-ranking Experimental Design}
\label{sec-spec-reranking-design}

We study whether generated specifications can improve function-level code generation via semantic re-ranking. For each problem, the model generates $n{=}10$ candidate solutions. A solution is considered correct if it passes all validated test cases for the problem, each consisting of a correct input--output pair obtained from the OJ system. As a baseline, we compute $\text{pass@}k$ by randomly selecting $k$ solutions from the $n$ candidates.

For specification-guided re-ranking, the model first generates a specification and five test inputs for each problem. Inputs violating the generated precondition are filtered out. The remaining inputs are executed on candidate solutions to obtain input--output pairs, which are checked against the generated postcondition. Solutions are ranked by their pass rate on these tests, and $\text{pass@}k$ is computed by selecting the top $k$ solutions after re-ranking.

\section{Impact of Prompt Length and Context Window}
\begin{figure}[tb]
    \centering
    \includegraphics[width=0.8\linewidth]{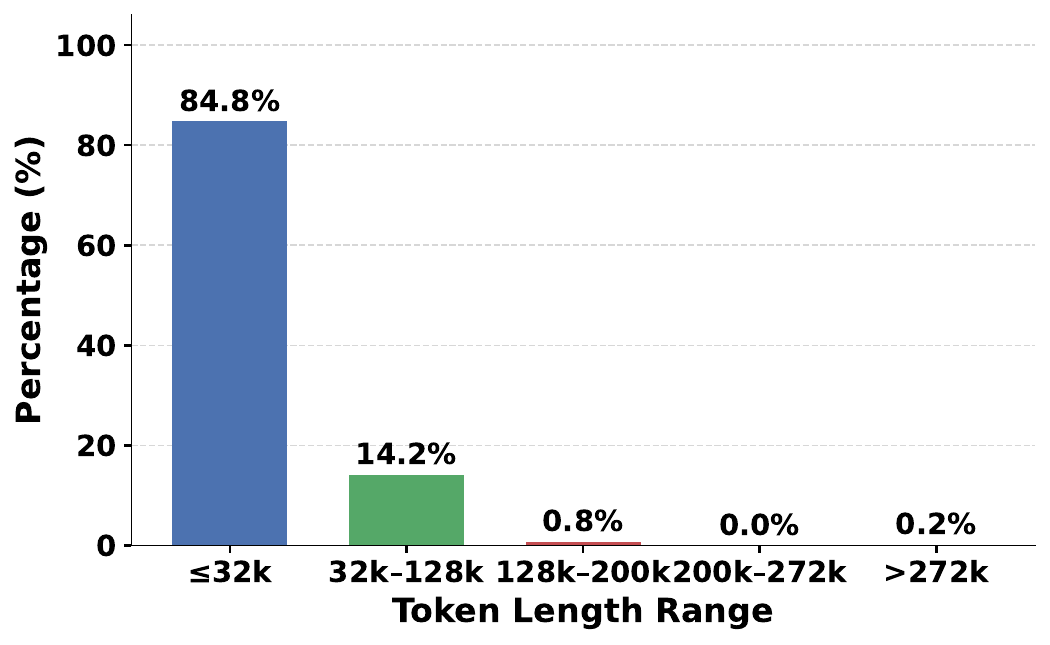}
    \caption{Distribution of prompt token lengths for \benchname{}-Repo, computed with the Qwen3 tokenizer. Each bar shows the proportion of inputs falling within a specific token length range.}
    \label{fig-tokens-distri}
\end{figure}

As shown in Figure~\ref{fig-tokens-distri}, the prompt length distribution is highly skewed toward short contexts: 84.8\% $\leq$ 32k, 14.2\% in 32k--128k, and only $\sim$1\% $\geq$ 128k.
Consequently, models with a 32k context window are disproportionately affected by the long tail of larger prompts, resulting in a noticeably lower pass rate.
In contrast, 128k-context models can accommodate nearly all inputs and exhibit more stable performance as prompt length increases.
Ultra-long-context models (200k--272k) mainly mitigate failures on a very small fraction of extreme cases, yielding only marginal improvements in overall averages.

\section{Postcondition Pass Rates Across Algorithmic Domains}

\begin{table*}[t]
\centering
\begin{tabular}{lc@{\hspace{10pt}}lc}
\toprule
\textbf{Algorithmic Domain} & \textbf{Pass Rate} &
\textbf{Algorithmic Domain} & \textbf{Pass Rate} \\
\midrule
Radix Sort & 66.7 & Geometry & 59.4 \\
Biconnected Component & 66.7 & Breadth-First Search & 59.3 \\
Bucket Sort & 66.7 & Monotonic Stack & 59.1 \\
Randomized & 66.7 & Backtracking & 59.1 \\
Suffix Array & 66.5 & Two Pointers & 59.0 \\
Merge Sort & 66.4 & Math & 58.9 \\
Binary Indexed Tree & 66.0 & Segment Tree & 58.9 \\
Counting Sort & 65.9 & Trie & 58.9 \\
Counting & 65.0 & Prefix Sum & 58.7 \\
Line Sweep & 64.5 & Dynamic Programming & 58.2 \\
Matrix & 63.7 & Brainteaser & 58.1 \\
Enumeration & 63.6 & Graph & 57.6 \\
String Matching & 63.4 & Topological Sort & 56.3 \\
Union Find & 62.6 & Memoization & 56.0 \\
Ordered Set & 62.4 & Divide and Conquer & 55.5 \\
Hash Table & 62.0 & Probability and Statistics & 55.5 \\
Game Theory & 61.6 & Depth-First Search & 55.4 \\
Stack & 61.4 & Sliding Window & 55.3 \\
Hash Function & 61.3 & Bitmask & 55.3 \\
Heap (Priority Queue) & 61.2 & Linked List & 54.2 \\
Array & 61.1 & Recursion & 51.9 \\
Shortest Path & 60.9 & Queue & 51.3 \\
Greedy & 60.8 & Tree & 47.9 \\
Sorting & 60.8 & Quickselect & 47.6 \\
Rolling Hash & 60.6 & Binary Tree & 46.2 \\
String & 60.5 & Eulerian Circuit & 44.0 \\
Monotonic Queue & 60.3 & Minimum Spanning Tree & 43.0 \\
Bit Manipulation & 60.0 & Combinatorics & 41.2 \\
Binary Search & 59.7 & Strongly Connected Component & 28.4 \\
Simulation & 59.6 & Binary Search Tree & 22.2 \\
Number Theory & 59.5 & & \\
\bottomrule
\end{tabular}
\caption{Postcondition pass rates (\%) for 61 algorithmic domains under GPT-5-mini on the original 2{,}494-task Func split.}
\label{tab-algorithmic-postcondition-passrate}
\end{table*}

Under the setting of GPT-5-mini, we analyze postcondition pass rates across algorithmic domains on the original 2,494-task Func split.
Each problem is associated with one or more of the 61 algorithmic tags provided by the LeetCodeDataset.
For each tag, we compute the average postcondition pass rate over all problems labeled with that tag.
The complete results are reported in Table~\ref{tab-algorithmic-postcondition-passrate}.

The statistics reveal a pronounced performance gradient across domains.
Sorting-related tasks (e.g., Radix Sort, Bucket Sort, and Merge Sort) achieve the highest pass rates (approximately 66\% on average).
In contrast, domains that require structural or recursive reasoning exhibit substantially lower pass rates.
For instance, Tree-related problems achieve 47.9\%, Binary Tree 46.2\%, Strongly Connected Component 28.4\%, and Binary Search Tree 22.2\%.

These results indicate that postcondition failures are systematically concentrated in domains where correctness depends on global structural properties or recursive invariants.
By contrast, tasks characterized by local and directly verifiable output properties (e.g., sorting correctness) are handled more reliably.

\section{Prompt Design}
\subsection{Prompt for Test Case Generation}
\label{sec-prompt-test-cases-gen}
\begin{prompt}
\begin{lstlisting}
You are given a coding problem described in natural language.
Your task is to write a Python code block containing **30 test case inputs** to evaluate whether the `preconditions(...)` function correctly implements the specification of the problem.

Here is the problem:

{PROBLEM_DESCRIPTION}

These test cases should be **carefully designed** to verify that the specification, i.e.:

* **Valid inputs** that satisfy the problem constraints.
* **Invalid inputs** that violate the constraints.

**Explicit restrictions:**

* Do **not** generate repetitive patterns.
* Use **only positional arguments**, do **not** use keyword arguments.
* Do **not** use excessively large inputs.
* Do **not** use operators such as `+`, `-`, `*`, `/`, `**`, etc.

Write the test cases in the format:

```python
preconditions(...)
```

Your output should be a single Python code block with exactly **30 test cases**, mixing both **valid and invalid inputs**, without any explanation or comments.
\end{lstlisting}
\end{prompt}

\subsection{Prompt for Solution Generation}
\label{sec-prompt-solution-gen}
\begin{prompt}
\begin{lstlisting}
You are given a coding problem described in natural language.
Your task is to implement a Python solution that adheres to the provided starter code.

**Problem Description**:

{PROBLEM_DESCRIPTION}

**Starter Code**:

```python
{STARTER_CODE}
```

Please write your solution within the following Python code block:

```python
# YOUR CODE HERE
```
\end{lstlisting}
\end{prompt}

\subsection{Prompt for Specification Generation}
\label{sec-prompt-specification-gen}
\subsubsection{Prompt for Specification Generation in \benchname{}-Func}

\begin{prompt}
\begin{lstlisting}
You are given a coding problem described in natural language.
Your task is to implement two Python functions: `preconditions` and `postconditions`.

* The `preconditions` function should **verify whether the given test input is valid** according to the problem constraints. Use only `assert` statements to check input validity.
* The `postconditions` function should **verify whether the given output is correct for the given input**, based on the expected behavior defined in the problem. Again, use only `assert` statements.

For both functions:

* Do **not** return any value.
* Raise `AssertionError` if any condition is violated.

Here is the problem:

{PROBLEM_DESCRIPTION}

Please implement two Python functions, `preconditions` and `postconditions`, **in the same Python code block**, and format your code as follows:

```python
# YOUR CODE HERE
```
\end{lstlisting}
\end{prompt}
The prompt provides explicit structural guidance. Specifically, it mandates fixed function names (\texttt{preconditions} and \texttt{postconditions}), requires the exclusive use of \texttt{assert} statements, forbids return values, and enforces a strictly defined output format limited to a single Python code block.
\subsubsection{Prompt for Specification Generation in \benchname{}-Repo}

\begin{prompt}
\begin{lstlisting}
You will be provided with a partial code base and an issue statement explaining a problem to resolve.
<issue>
{ISSUE_STATEMENT}
</issue>
<code>
{REPOSITORY_FILES}
</code>

Please write a conftest.py file that defines pytest_sessionstart, along with precondition and postcondition functions for this issue. You should import the target function that requires modification, along with any other necessary libraries, then use the pytest_sessionstart hook to wrap this target function. This ensures that when pytest runs, the patch automatically triggers both the precondition and postcondition functions. The precondition and postcondition should verify whether the target function's behavior aligns with the issue statement after modification. 

Hints:
File: <modified file name>
  MODIFY:
    - <modified function name>

Respond with the complete conftest.py implementation in the following format.

<conftest.py>
import pytest
import chat.utils as utils

def pytest_sessionstart(session):
    original_add = utils.add

    def wrapped_add(num1, num2):
        precondition(num1, num2)
        result = original_add(num1, num2)
        postcondition(num1, num2, result)
        return result

    utils.add = wrapped_add
    
def precondition(num1, num2):
    assert isinstance(num1, (int, float)), "num1 must be int or float"
    assert isinstance(num2, (int, float)), "num2 must be int or float"
    return True

def postcondition(num1, num2, result):
    assert result == num1 + num2, "result must equal num1 + num2"
    return True
</conftest.py>

\end{lstlisting}
\end{prompt}
The prompt includes a description of the issue, the relevant repository files associated with the issue, and the name of the function that needs to be modified. It also provides an example of the expected model output.
The relevant repository files are provided by the SWE-bench Verified dataset under its Oracle setting, which includes the exact files modified by the ground-truth patch that resolved the issue on GitHub. All retrieved files, including original library imports and repository code, are included inside the \texttt{<code> ... </code>} section of the prompt.

\section{Model Deployment Details}
\label{sec-model-deployment-details}
\begin{table}[!htbp]
\centering
\scriptsize
\setlength{\tabcolsep}{2.5pt}
\begin{tabular}{l c c c c c}
\toprule
\textbf{Model} & \textbf{Reason.} & \textbf{Temp.} & \textbf{TopP} & \textbf{TopK} & \textbf{MinP} \\
\midrule
Qwen3 (0.6B--32B) & \ding{55}  & 0.7 & 0.8  & 20 & 0 \\
Qwen3 (0.6B--32B) & \ding{51}      & 0.6 & 0.95 & 20 & 0 \\
QWQ-32B           & \ding{51}            & 0.6 & 0.95 & 20 & 0 \\
GPT-OSS (20B--120B) & \ding{51}          & 1.0 & 1.0  & -- & -- \\
DeepSeek-V3.2     & \ding{55}            & 1.0 & 0.95 & -- & -- \\
\bottomrule
\end{tabular}
\caption{Sampling parameters for different models}
\label{tab-sampling-params}
\end{table}

Our evaluation includes Qwen3 (0.6B, 1.7B, 4B, 8B, 14B, 32B)~\cite{yang2025qwen3technicalreport}, QWQ-32B~\cite{qwq32b}, DeepSeek-V3.2~\cite{deepseekai2025deepseekv32pushingfrontieropen}, GPT-OSS 20B/120B~\cite{openai2025gptoss120bgptoss20bmodel}, Claude-4.5-Sonnet~\cite{claude45_sonnet}, GPT-5-mini, GPT-5~\cite{openai_gpt5}, Gemini-2.5-Flash, and Gemini-2.5-Pro~\cite{google_gemini_2.5_2025}. This suite covers open-source instruction models, reasoning-enhanced variants, and frontier proprietary systems.

For deployment, Qwen3 (0.6B--32B), QWQ-32B, and GPT-OSS 20B are executed locally using \texttt{vLLM} on eight NVIDIA H200 GPUs; all remaining models are queried through their official APIs.
For specification generation, we use the inference settings in Table~\ref{tab-sampling-params} and report $\text{pass@}1$~\cite{chen2021evaluating} from a single generation per prompt.

\section{Issue-Only Exposure Probe}
\label{sec-repo-exposure-probe}
\benchname{}-Repo draws on public SWE-bench Verified issues, so models may have encountered related issues or patches during training. We gave GPT-5-mini only the repository name and issue statement and asked it to predict which files and functions or classes a patch would modify. We flagged an issue if the model predicted at least one exact reference file path and at least one changed function or class, and the issue statement contained none of the reference file paths, basenames, stems, or changed symbol names. We extracted changed function or class names from the reference patch hunk headers for 494 of the 500 issues. Of these, 34 met the flagging criteria.

\begin{table*}[tb]
\centering
\small
\begin{tabular}{lccc}
\toprule
\textbf{Model} & \textbf{Flagged (34)} & \textbf{Remaining (466)} & \textbf{Full (500)} \\
\midrule
Claude-4.5-Sonnet & 6/34 (17.6\%) & 95/466 (20.4\%) & 101/500 (20.2\%) \\
Gemini-2.5-Pro & 7/34 (20.6\%) & 84/466 (18.0\%) & 91/500 (18.2\%) \\
GPT-5-mini & 4/34 (11.8\%) & 44/466 (9.4\%) & 48/500 (9.6\%) \\
GPT-5 & 3/34 (8.8\%) & 40/466 (8.6\%) & 43/500 (8.6\%) \\
DeepSeek-V3.2 & 6/34 (17.6\%) & 28/466 (6.0\%) & 34/500 (6.8\%) \\
\bottomrule
\end{tabular}
\caption{Strict specification pass counts and rates on \benchname{}-Repo. The remaining issues include six for which a changed function or class name could not be extracted.}
\label{tab-repo-exposure-probe}
\end{table*}

Table~\ref{tab-repo-exposure-probe} summarizes strict pass from the existing repository-level specification runs. Among the models shown, the highest strict pass rate on the flagged issues is 20.6\% (7/34). Correct location predictions can come from prior exposure, familiarity with repository conventions, or inference from the issue text.

\end{document}